\documentclass[12pt]{article}

\RequirePackage[numbers]{natbib}
\usepackage{array}
 \usepackage{amsmath,amssymb}
 \usepackage{amsthm}
 \usepackage{bm}
 \usepackage{latexsym}
 \usepackage{CJK}
 \usepackage{multirow}
 \usepackage{caption}
 \usepackage{graphicx,color}
 \usepackage{amssymb,amsfonts}
 \usepackage{hyperref}
 \usepackage[english]{babel}
 \usepackage{times}
 \usepackage{enumerate}
\usepackage{geometry}
\usepackage{float}
\usepackage{booktabs}
\usepackage{endnotes}
\usepackage{mathtools}
\usepackage{amscd}
\usepackage{bbm}
\usepackage{multirow}
\usepackage{booktabs}
\usepackage{graphicx}
\usepackage{adjustbox}

\usepackage[scaled]{beramono}
\usepackage[T1]{fontenc}
\usepackage{amsmath,amsfonts,amssymb}
\usepackage{mathtools}
\usepackage{url}
\usepackage{hyperref}
\usepackage{setspace}

\usepackage{fixltx2e}
\usepackage[dvipsnames]{xcolor}

\usepackage[toc]{appendix}

\usepackage[utf8]{inputenc}
\usepackage{authblk}

\usepackage{booktabs}
\usepackage{ctable}
\usepackage{changepage}
\usepackage{subfig}

\newtheorem{innercustomgeneric}{\customgenericname}
\providecommand{\customgenericname}{}
\newcommand{\newcustomtheorem}[2]{%
  \newenvironment{#1}[1]
  {%
   \renewcommand\customgenericname{#2}%
   \renewcommand\theinnercustomgeneric{##1}%
   \innercustomgeneric
  }
  {\endinnercustomgeneric}
}

\newcustomtheorem{customlemma}{lemma}

\newtheorem{theorem}{Theorem}[section]

\DeclareMathOperator*{\argmin}{argmin}

\begin{document}

		\title{High-order Joint Embedding for Multi-Level Link Prediction}
		\author{Yubai Yuan and Annie Qu\footnote{Yubai Yuan is Postdoctoral Researcher, Department of Statistics, University of California, Irvine (E-mail: yubaiy@uci.edu). Annie Qu is Chancellor's Professor, Department of Statistics, University of California, Irvine (E-mail: aqu2@uci.edu). This work is supported by NSF Grants DMS 1952406.
		}}
		\date{ }
		\maketitle
		\vspace{-4mm}
		\begin{abstract}
Link prediction infers potential links 
from observed networks, and is one of the essential problems in network analyses. In contrast to traditional graph representation modeling which only predicts two-way pairwise relations, we propose a novel tensor-based joint network embedding approach on simultaneously encoding pairwise links and hyperlinks onto a latent space, which captures the dependency between pairwise and multi-way links in inferring potential unobserved hyperlinks. The major advantage of the proposed embedding procedure is that it incorporates both the pairwise relationships and subgroup-wise structure among nodes to capture richer network information. In addition, the proposed method introduces a hierarchical dependency among links to infer potential hyperlinks, and leads to better link prediction. In theory we establish the estimation consistency for the proposed embedding approach, and provide a faster convergence rate compared to link prediction utilizing pairwise links or hyperlinks only. Numerical studies on both simulation settings and Facebook ego-networks indicate that the proposed method improves both hyperlink and pairwise link prediction accuracy compared to existing link prediction algorithms.

\noindent\textbf{Key words:} data augmentation, hypergraph, latent factor model, method of moments, non-convex optimization, shared parameters, symmetric tensor completion.
		\end{abstract}

\section{Introduction}

Hyperlinks or hyperedges 
generalize traditional pairwise links through capturing interactions among groups of nodes. For example, hyperlinks occur frequently in online social networks and recommender systems 
such as Delicious, Last.fm and Flickr, where the system involves not only pairwise relations between users and items, but also three-way 
user-tag-item relations that cannot be captured by pairwise relations using the traditional network 
formulation.  
\color{black}
{Another example of a hyperlink is in a gene interaction network, where it is essential to identify subgroups of genes which are functionally associated with 
each other to potentially formulate a  protein complex \citep{das2012hint, razick2008irefindex,croft2013reactome,
fabregat2015reactome, shojaie2009analysis, rinaldo2005characterization}.  
In contrast, pairwise relations between genes are inadequate to capture the collaborative high-order interactions of one gene
with a subgroup of other genes. 
\begin{figure}[H]
\centering 
   \includegraphics[width=4.0in,height=1.6in]{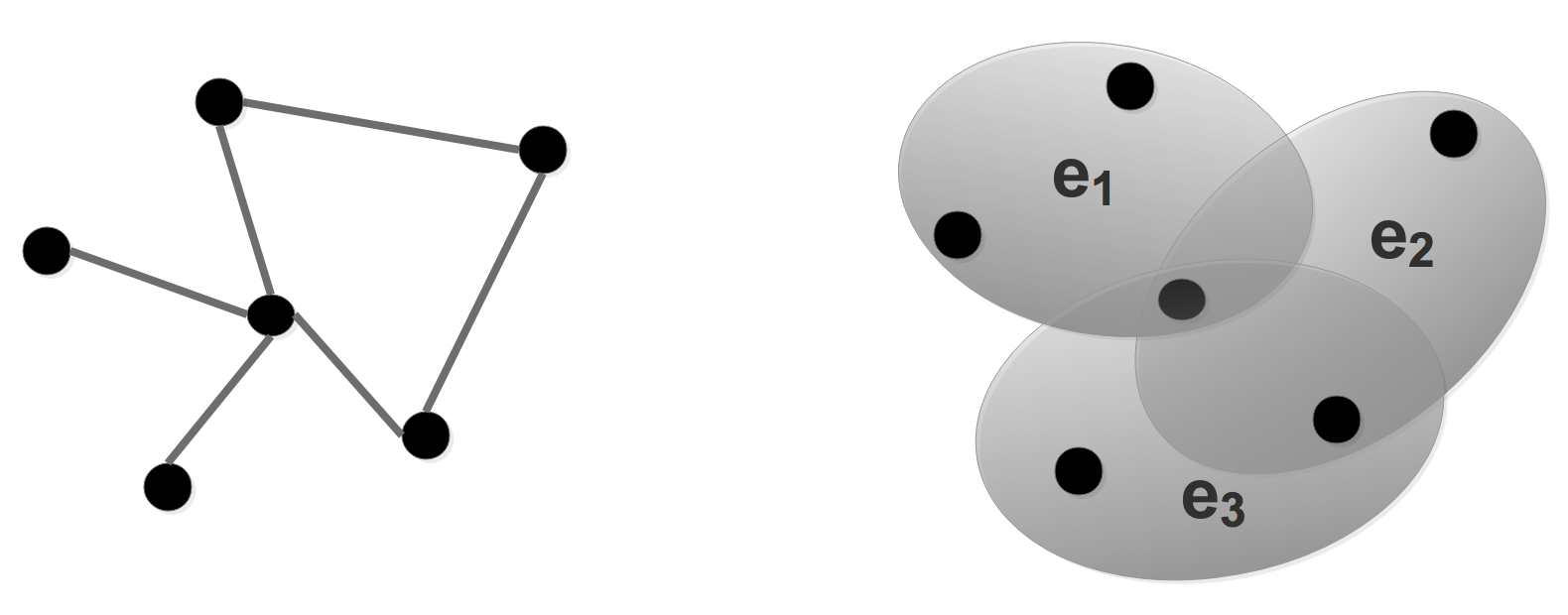}
\caption{The left network illustrates pairwise links. The right figure illustrates three hyperlinks $e_1, e_2$ and $e_3$, where each of them is a 3-order hyperlinks connecting 3 nodes.}
\label{fig:Fig1}
\end{figure}
{A hyperlink is called $m$-order if it is a set containing $m$ nodes. 
Figure \ref{fig:Fig1}  illustrates  the differences  between pairwise links and hyperlinks.} 
{Motivated by previous examples, it is critical to identify potential multi-way or high-order relations among
multiple nodes represented by hyperlinks.} Meanwhile, the coexistence of hyperlinks and pairwise links within a complex network system also motivates us to model and utilize their relations simultaneously for inference. In contrast, 
existing link prediction methods are designed for inferring either two-way or multi-way relations separately, which ignore the joint information between pairwise links and hyperlinks within the same network. For example, most existing methods utilize the information only from observed two-way relations \citep{hristova2016multilayer,jalili2017link,song2009scalable,al2011survey, menon2011link,liben2007link,lu2011link,
al2006link, zhao2017link, 10.1093/biomet/asx042} or decompose multi-way relations into pairwise relations \citep{cao2015grarep,tang2015line,grover2016node2vec}. In general, these methods are unable to capture high-order relations among nodes and suffer from information loss in the process of decomposing multi-way relations to two-way relations.   
 
Alternatively, hyperlink prediction methods are proposed through directly modeling observed multi-way relations \citep{klamt2009hypergraphs,li2013link,zlatic2009hypergraph,agarwal2006higher, zhou2007learning,zhu2016heterogeneous,ghoshal2009random}. 
One unique challenge for these methods is that hyperlinks are only partially observed or highly sparse
in practice, which is quite distinct from the pairwise relations which are likely to have more observed complete information. 
For example, identification of a hyperlink for  
a protein complex connecting multiple genes 
 requires additional biological experiments and validation, and in social networks, inferences about local social circles may require additional information involving high-order interactions among users.
Therefore, inferring high-order relations with a limited training size could be unreliable without borrowing information from observed pairwise links.   
}

\color{black}
{
Existing works on hyperlink modeling are also considered in
node classification and community detection. For instance, \citep{chitra2019random,pu2012hypergraph, agarwal2006higher} 
suggest methods based on hyperlink expansions or 
random walks to reconstruct hyperlinks from pairwise links. These methods use a principle of generating
hyperlinks based on pre-specified relations among 
hyperlinks and pairwise links, while treating hyperlinks as a subgraph with a certain configuration 
such as a fully-connected clique or star-shaped subgroup of nodes. However, this heuristic principle 
is not adaptive to different structures of hyperlinks, which tend to lead 
misspecified hyperlinks, especially in the absence of the accurate knowledge of the pairwise links and hyperlinks.  
In addition,  they mainly focus on node classifications and 
detection of global community structures rather than  identifying subgroup structures.}

To overcome the foregoing difficulties of modeling relations with heterogeneous orders, in this paper we develop a novel network embedding procedure to jointly model pairwise links and high-order 
hyperlinks simultaneously to capture complex interactions among nodes. In particular, we proceed in a hierarchical
fashion in that pairwise and multiway relations are modeled at different resolutions. E.g.,
pairwise relations are structured by low-level node-wise network features, while  
hyperlinks  capture multiway relations via  high-level subgroup-wise features. Jointly they can identify the subgroup configuration and capture the hyperlink-generating features effectively. This enables the incorporation of more complete 
and rich information from observed networks. 

One advantage of this hierarchical modeling is the mutual information borrowing between hyperlink prediction and pairwise link prediction.
More specifically, in the presence of a hyperlink, nodes from the same hyperlink are more likely to form a pairwise 
relation as compared with nodes in the absence of a hyperlink. On the other hand, nodes that are
highly connected by pairwise links may suggest the presence of a potential hyperlink among them. In addition, the proposed joint embedding framework follows a principle of data augmentation for hyperlinks based on pairwise links, which improves inference accuracy as the effective sample size of links increases, and consequently 
captures subgroup structures associated 
with potential hyperlinks more efficiently. Furthermore, this principle of network connectivity also reflects the phenomenon in practice.   
In summary, the proposed hyperlink prediction framework provides cohesive statistical modeling for both 
pairwise links and hyperlinks, which enhance the mutual inferences between pairwise links and hyperlinks. 

\color{black}
{In theory, we provide the consistency of the proposed estimation and show that our method achieves a faster convergence rate compared to embedding procedures utilizing pairwise link information only, since we are able to incorporate at least one of the observed or inferred hyperlinks via the joint embedding procedure. The results are valid in cases when either pairwise link and hyperlink are independent or dependent conditioning on the latent space. In addition, our theoretical analyses show that hyperlink augmentation can further boost the convergence rate of the joint embedding method, especially when the conditional dependency among pairwise and hyperlinks is strong.} The theoretical development 
using large deviation theory is nontrivial here, since pairwise links and hyperlinks are correlated intrinsically in that the independent model assumption cannot be utilized. In contrast, most existing probability concentration properties are established under the independent model assumption. 
 
\color{black}
This paper is organized as follows: Section \ref{2.1} introduces the background and notation of the proposed method. Section \ref{2.2} introduces the proposed joint embedding method and the inference procedure from observed pairwise network to hyperlinks. Section \ref{2.3} establishes the theoretical properties of the proposed embedding method under the  scenarios where the hyperlinks are either observed or inferred from pairwise links. Section \ref{2.4} demonstrates simulation studies, and Section \ref{2.5} illustrates an application of the Facebook ego-network. The last section provides conclusions and some final discussion.

\section{Background and Notations}\label{2.1}
We define  an observed  network $\bm{G} = (\bm{V}, \bm{E})$,  
where $\bm{V} = \{v_i\}_{i=1}^N$ denotes a set of $N$ nodes 
and {$\bm{E} \subset {\bm{V} \choose 2} $ is a set of node pairs with observed pairwise link status, 
which represents the presence or absence of a pairwise link.} For an undirected and unweighted network, 
$\bm{G}$ can be represented through a symmetric binary adjacent matrix 
$\bm{Y} = \{Y_{ij}\}_{1\leq i\neq j \leq N}$, {in that 
\begin{align*}
Y_{ij} = \begin{cases*} 
    1,  \; \{i,j\}\in \bm{E}\; \text{and} (i,j) \; \text{is presence},  \\
    0,  \; \{i,j\}\in \bm{E}\; \text{and} (i,j) \; \text{is absence},\\
   -1, \; \{i,j\}\notin \bm{E}.
    \end{cases*}
\end{align*}}
{In addition, we define an $m$-order uniform hypergraph on a set of nodes $\bm{V}$ as $\bm{G}_{\bm{H}} = \{\bm{V}, \bm{H}\subset {\bm{V} \choose m}\}$, where $e\in \bm{H}$ is an index set of $m$-tuple indices $e = \{i_1, i_2,\cdots, i_m\}$ with observed hyperlink status, indicating the presence or absence of a hyperlink. To represent the $m$-order hyperlink status, we introduce an $m$-order tensor $\mathcal{Y}\in \{0,1,-1\}^{R^m}$ such that
\begin{align*}
Y_{i_1i_2\cdots i_m} = \begin{cases*} 
    1,  \; \{i_1,i_2\cdots i_m\}\in \bm{H}\; \text{and} \{i_1,i_2\cdots i_m\} \; \text{is presence},  \\
    0,  \; \{i_1,i_2\cdots i_m\}\in \bm{H}\; \text{and} \{i_1,i_2\cdots i_m\} \; \text{is absence},\\
   -1, \; \{i_1,i_2\cdots i_m\}\notin \bm{H}.
    \end{cases*}
\end{align*}}

{We introduce observation indicator $\bm{A} = \big\{A_{ij}\in \{0,1\}, A_{i_1\cdots i_m}\in\{0,1\}\big\}$, where $A_{ij}=1$ if $\{i,j\}\in \bm{E}$ and $A_{ij}=0$ if $\{i,j\}\notin \bm{E}$, and $\{A_{i_1\cdots i_m}\}$ are similarly defined. Given that $\bm{A}$ is independent from link statuses, the network modeling $P(\bm{Y},\bm{\mathcal{Y}})$ can be decomposed into two submodels, where one models the underlying link statuses: $P(\bm{Y}\in\{0,1\},\bm{\mathcal{Y}}\in\{0,1\})$, and the other models whether a link status is observed or not: $P(\bm{A}\in\{0,1\})$. In the following, we focus on modeling the link statuses based on observed link statuses where, $\{Y_{ij}\}\in\{0,1\}$ and $\{Y_{i_1 \cdots i_m}\} \in\{0,1\}$. For the observation modeling $P(\bm{A}\in\{0,1\})$, we follow the convention in literature \citep{huisman2009imputation} and assume the observed link statuses are uniformly and randomly sampled from all possible links on networks, i.e., $\{A_{ij}, A_{i_1\cdots i_m}\}$ are independent binary random variables. However, we also consider the scenario of observed link statuses with missing not at random in that dependency exists among $\{A_{ij}\}$, and provide the theoretical and numerical analysis in later sections.}



Denote the sets of observed pairwise link statuses and hyperlink statuses as $\Omega_{\bm{Y}}$ and $\Omega_{\mathcal{Y}}$. Accordingly, the number of observed pairwise link statuses and hyperlink statuses are $|\Omega_{\bm{Y}}|$ and $|\Omega_{\mathcal{Y}}|$. 
The proposed link prediction framework is based on network latent space modeling. Specifically, we introduce an $r$-dimensional latent vector representation $Z_i = (Z_{i1}, \cdots,Z_{ir})$ for each node $v_i$, where each element $Z_{ik},\; k=1,\dots,r$ represents a latent feature of node $v_i$. The latent factor modeling enables us to capture the relation between nodes $v_i$ and $v_j$ on a network through their concordance on the latent space spanned by $\{Z_i\}_{i=1}^N$. In addition, the latent factor model allows us to capture the low-rank structure feature by assuming that the dimension of the latent space $r$ is much smaller than $N$. 

Through a dimension reduction we are able to increase the estimation efficiency of latent factors while reducing the variation of prediction. Based on the node-wise latent factors $\{Z_i\}_{i=1}^N$,  we propose a probabilistic link modeling. That is, we assume that both pairwise link $Y_{ij}$ and hyperlink $Y_{i_1i_2\cdots i_m}$ follow a Bernoulli distribution with probability $\bm{P}\big(Y_{ij}|Z_i,Z_j\big)$ and $\bm{P}\big(Y_{i_1\cdots i_m}|Z_{i_1},\!\cdots,\!Z_{i_m}\big)$, respectively. Therefore, once the positions of nodes in the latent space are estimated, we can predict the probabilities for both pairwise links and hyperlinks among these nodes.


\vspace{-2mm}

\section{Methodology}\label{2.2}

In this section, we introduce the proposed joint network embedding framework to perform prediction for both pairwise links and hyperlinks. The key idea is to project the nodes in the network into points in a latent space, and preserve both the observed pairwise and high-order relations in the latent space. In the following, we present the methods for embedding pairwise links and hyperlinks.
\vspace{-6mm}
\subsection{Reconstruction of Pairwise Relations in Latent Space}  
To incorporate observed pairwise relations into a latent space, we propose to minimize a loss function to estimate node-wise latent factors $\bm{Z} = \{Z_i \in \mathbb{R}^r\}_{i=1}^N$:
\begin{align}\label{eq1}
L_1(\bm{Z}) = 
\frac{1}{|\Omega_{\bm{Y}}|}\sum_{Y_{ij}\in \Omega_{\bm{Y}}}(Y_{ij} - \sigma\Big[Z_j^TZ_i \Big])^2,
\end{align}
where $|\Omega_{\bm{Y}}|$ is the number of observed pairwise links, and {the $Z_{j}^TZ_{i}$ is the second-order concordance, measuring the degree of similarity between two nodes' latent features.} The $\sigma(\cdot)$ is a logistic link function that transforms the second-order concordance into the probability of a pairwise link. Intuitively, two latent factors $Z_i$ and $Z_j$ are expected to be close in the latent space given that a pairwise link exists between nodes $v_i$ and $v_j$, i.e., $Y_{ij}= 1$. In contrast, two latent factors $Z_i$ and $Z_j$ are far away from each other given $Y_{ij} = 0$. In this way, we transform the binary link statuses of the network into continuous measures of distance in the latent space. 

The probabilistic prediction loss in (\ref{eq1}) has the same form as the Brier score \citep{brier1950verification}, e.g., $(Y-\hat{P})^2$ where $Y$ is a binary outcome and $\hat{P}$ is the corresponding predicted probability. Compared with the negative log-likelihood loss, e.g., $-[Y\log\hat{P} + (1-Y)\log (1-\hat{P})]$, both of them have a similar U-shape landscape and reach the same minimizer of $\hat{P} = P(Y=1)$ at the population level. In addition, an empirical 
study \citep{menon2011link} demonstrates that the choice of loss function has minor influence 
on link prediction performance using latent factor models. 

\vspace{-2mm}
\subsection{Reconstruction of High-order Relations in Latent Space} 
Hyperlink prediction is more challenging since the configurations of high-order relation involve a high degree of uncertainty and require high-dimensional modeling tools. 
One key element of the  proposed method is to model hyperlinks via the structure of latent factors 
expressed in high-order tensors. 
We formulate the $m$-order hyperlinks by an $m$-order tensor $\mathcal{Y}=(Y_{i_1i_2\cdots i_m})\in \{0,1\}^{N^m}$. Specifically, if there is a hyperlink connecting $m$ nodes $\{v_{i_1},\cdots, v_{i_m}\}$, then $Y_{i_1i_2\cdots i_m} =1$. To map the high-order relations of the network into the latent space, we model the $m$-order hyperlinks as a low-rank concordance structure on the latent feature $\bm{Z}$ based on the CANDECOMP/PARAFAC (CP) tensor decomposition. Specifically, we introduce a high-order concordance measure function among $m$ nodes as 
\vspace{-3mm}
\begin{align}\label{high-order}
f(Z_{i_1},Z_{i_2},\cdots,Z_{i_m}) = \sum_{k=1}^r\psi_{k,i_1i_2,\cdots i_m} |Z_{i_1k}Z_{i_2k}\cdots Z_{i_mk}|,
\end{align}
where 
$$\psi_{k,i_1i_2,\cdots i_m} = \begin{cases*} 
    1  \;\; \text{$Z_{i_lk}\geq 0$, $l = 1,\cdots,m$\; or \; $Z_{i_lk}< 0$, $l = 1,\cdots,m$},   \\
    -1 \;\; \text{otherwise}.
    \end{cases*}   $$
{The high-order concordance measures the degree of similarity among a group of nodes.} Here we constrain the scope of concordance among multiple latent features only when their signs are all the same, thereby by increasing the power for discriminating between high-order concordance and high-order discordance. Based on the high-order concordance among the latent factors of nodes $v_{i_1},v_{i_2},\cdots, v_{i_m}$, we model the probability of hyperlink $Y_{i_1i_2\cdots i_m}$ as:  
\begin{align}\label{hyper_formula}
P(Y_{i_1i_2\cdots i_m} = 1) = \sigma(\sum_{(i,j)\in\{i_1i_2,\cdots,i_m\}}Z_i^TZ_j + \beta f(Z_{i_1},Z_{i_2},\cdots,Z_{i_m})),
\end{align}
where $\beta>1$ is a weighting parameter. Analogous to measuring the pairwise concordance via an inner product among latent factors, we apply the generalized inner product $\sum_{k=1}^r\psi_kZ_{i_1k}Z_{i_2k}\cdots Z_{i_mk}$ to measure the joint concordance among a group of $m$ latent factors. Notice that the proposed measure $f(Z_{i_1},Z_{i_2},\cdots,Z_{i_m})$ is equivalent to the CP tensor decomposition given that $\psi_k = 1, k = 1,\cdots,r$. The purpose of introducing sign consistency function $\psi_k$ for each latent factor dimension $k = 1,\cdots,r$ is to adjust the sign of multiplication among latent factors so that the sign is consistent with the interpretation of $f(Z_{i_1},Z_{i_2},\cdots,Z_{i_m})$. For example, consider the $k$th latent feature for three nodes $Z_{i_1k}>0,Z_{i_2k}<0, Z_{i_3k}<0$. Since the signs of the latent features are not consistent, the $P(Y_{i_1i_2i_3} = 1)$  decreases given other latent features. However, applying CP decomposition would lead to $Z_{i_1k}Z_{i_2k}Z_{i_3k}>0$, which increases the $P(Y_{i_1i_2i_3} = 1)$.  Therefore, the introduction of $\psi_k, k = 1,\cdots,r$ encourages a large value of $f(Z_{i_1},Z_{i_2},\cdots,Z_{i_m})$ only when all $Z_{i_1},Z_{i_2},\cdots,Z_{i_m}$ are close to each other in the latent space, which is consistent with the interpretation of the high-order concordance. Note that the joint  concordance cannot be  directly inferred by the pairwise concordance in the sense that even $Z_iZ_j^T$ is large for any pair $(i,j)\in \{i_1,i_2,\cdots,i_m\}$,  though it is still  possible that  $f(Z_{i_1},Z_{i_2},\cdots,Z_{i_m})$ is small. This also implies that the joint concordance capturing high-order relations  cannot be substituted for by pairwise concordance.

The latent-factor-based hyperlink modeling in (\ref{hyper_formula}) introduces a hierarchical dependency between pairwise links and hyperlinks since (\ref{hyper_formula}) also incorporates the pairwise concordance among nodes in (\ref{eq1}). In addition, the high-order concordance is correlated with $\sum_{(i,j)\in\{i_1i_2,\cdots,i_m\}}Z_i^TZ_j$ via the shared latent factors. Through utilizing  
the dependency between hyperlink and pairwise link, we borrow information from each for more accurate prediction. In addition, the proposed hierarchical dependency facilitates  better interpretations in real applications as compared to the methods of using pairwise links or hyperlinks separately. 


Subsequently, we incorporate the hyperlink information into the latent space of nodes via obtaining the latent factors $\bm{Z}$ via minimizing the following hyperlink loss function:
\vspace*{-2mm}
\begin{align}\label{eq2}
\vspace*{-2mm}
L_2(\bm{Z})\! =\! 
\frac{1}{|\Omega_{\mathcal{Y}}|}\!\sum_{Y_{i_1i_2 \cdots i_m}\in \Omega_{\mathcal{Y}}}\!\!\!\!\!w_{i_1i_2\cdots i_m}\Big\{Y_{i_1i_2\cdots i_m}\!\!-\sigma\big(\!\!\!\sum_{(i,j)\in\{i_1i_2,\cdots,i_m\}}\!\!\!\!\!\!\!\!Z_i^TZ_j\!+\!\beta f(Z_{i_1},Z_{i_2},\cdots,Z_{i_m}) \big )\Big\}^2,
\vspace*{-2mm}
\end{align}
where $w_{i_1i_2,\cdots,i_m}$ is the weight for hyperlink $Y_{i_1i_2\cdots i_m}$,  $\Omega_{\mathcal{Y}}$ is the set of incorporated $m$-order hyperlinks, and $|\Omega_{\mathcal{Y}}|$ is the total number of hyperlinks. If there exists a potential hyperlink connecting nodes $\{i_1, i_2, \cdots, i_m\}$  in a hypergraph, then decreasing the loss function in (\ref{eq2}) leads to a strong joint concordance among latent factors $\{Z_{i_1}, Z_{i_2}, \cdots, Z_{i_m}\}$. Consequently, we preserve high-order relations among nodes associated with their embedding in latent space. In terms of latent factors estimation, \eqref{eq1} and \eqref{eq2} serve as the second-order moments and the $m$-order moment estimations of $\bm{Z}$, respectively. 
Intuitively, incorporating additional moment information generated from latent factors $\bm{Z}$ reduces estimation bias while increasing efficiency, which leads to a more accurate estimation of latent factor $\bm{Z}$.
\subsection{Joint Network Embedding for Pairwise Link and Hyperlink Prediction} 
Based on the link embedding (\ref{eq1}) and (\ref{eq2}), we estimate the latent features of nodes by jointly incorporating pairwise links and hyperlinks. Due to randomness or noisy sources of information, some pairwise links and hyperlinks can be incorrectly observed or inferred, which requires regularization during the embedding process. In addition, in a network system with many potential hyperlinks, we are interested in detecting more significant hyperlinks in the sense that they capture local subgroup structures with high certainty in the network. To achieve these goals, we 
propose the following loss function:
\begin{align}\label{eq3}
L(\bm{Z}) &= \frac{1}{|\Omega_{\mathcal{Y}}|} \!\sum_{Y_{i_1\cdots i_m}\in \Omega_{\mathcal{Y}} }\!\!\!\!\!w_{i_1\cdots i_m}\Big\{Y_{i_1\cdots i_m}\!-\!\sigma\big(\!\!\sum_{(i,j)\in\{i_1,\cdots,i_m\}}\!\!\!Z_i^TZ_j+\beta f(Z_{i_1},\cdots,Z_{i_m})\Big\}^2\\ \nonumber
& + \frac{1}{|\Omega_{\bm{Y}}|}\sum_{Y_{ij}\in \Omega_{\bm{Y}}}(Y_{ij} - \sigma\Big[Z_i^TZ_j\Big] )^2 
\!+\! \lambda\|\bm{Z}\|^2. \nonumber
\end{align}
Specifically, we impose the weight function $w_{i_1i_2\cdots i_m}$ to downweigh spurious links through penalization. We also adopt the penalization $\|\bm{Z}\|^2$ to control the magnitude of the latent factors of nodes in order to alleviate the overfitting of latent factors due to spurious pairwise links and hyperlinks. Furthermore, the penalization $\|\bm{Z}\|^2$ imposes a low-rank structure of latent factors which can mitigate the singularity problem when the degree of nodes is smaller than the rank of latent factors. 
{To tune the penalty parameter $\lambda$ in (\ref{eq3}), we first split both the observed pairwise link status and hyperlink status into a training and validation dataset. Then we fit the loss function (\ref{eq3}) on the training dataset and perform a grid search for $\lambda$ on a predefined region, e.g., $[0,1]$. We can select a $\lambda$ to maximize the AUC on the validation dataset.}



After mapping each node into the latent space spanned by 
column vectors of $\bm{Z}$ consisting of the pairwise links and hyperlinks, we predict potential pairwise links and hyperlinks through an estimated degree of concordance among the latent factors of nodes. Specifically, we predict a pairwise link between nodes $v_i$ and $v_j$ through
\begin{align} \label{pre1}
P\big(Y_{ij}= 1|(v_i, v_j)\big) = \sigma (Z_i^TZ_j). 
\end{align} 
Similarly, to predict an $m$-order hyperlink among a group of nodes
$v_{i_1}, v_{i_2}, \cdots, v_{i_m}$, we have
\begin{align} \label{pre2}
P\big(Y_{i_1i_2\cdots i_m}= 1|(v_{i_1}, v_{i_2}, \cdots, v_{i_m})\big) = \sigma\Big[\sum_{(i,j)\in\{i_1i_2,\cdots,i_m\}}Z_i^TZ_j+\beta f(Z_{i_1},Z_{i_2},\cdots,Z_{i_m})\Big].
\end{align} 
Although the main focus in this paper is link prediction, identifying the latent factor space of nodes is also fundamentally
important  as it permits exploration of other types of network structures. For example, in community detection,  
detection of the community structures of a network leads to  discoveries in biology, marketing, and social science. 
In other  situations, we may develop a clustering algorithm to identify homogeneous subgroups
of nodes in a latent space based on their estimated similar
embeddings of learned nodes. 

\noindent{\textbf{Remark}: The proposed joint embedding procedure can be generalized to the case when orders of hyperlink statuses are different. Instead of directly encoding hyperlink statuses via
a series of tensors with different orders, an alternative solution is to decompose hyperlink statuses of different orders into hyperlink statuses with the same and low order, which could be then formulated into a single hyperlink tensor. In this way, an original high-order hyperlink statuses is approximated by a set of low-order hyperlink statuses, and then the objective function (\ref{eq3}) can be directly applied.}

%

\subsection{Inferring Potential Hyperlinks through Observed Links} 

An innovation of the proposed method is that the framework enables a data augmentation procedure on hyperlinks based on the observed network. Although in many applications, only a small number of high-order relations represented by hyperlinks can be directly observed, it is feasible to infer potential hyperlinks from observed links through link dependency information. 

The rationale is that pairwise links and hyperlinks characterize the same types of relations within a network but at different group levels. Therefore, observing whether two nodes are bilaterally connected or not is informative for inferring whether they belong to the same subgroup or hyperlink. This hierarchical dependency among links is implicitly introduced into the proposed framework as the node-wise latent factors and their concordances are shared by both pairwise link (\ref{eq1}) and hyperlink modeling (\ref{hyper_formula}), which enables us to infer the unobserved hyperlinks along with estimating latent factors. The inferred hyperlinks can be integrated into the embedding procedure to increase the effective sample size and enrich the subgroup structures in the latent space.     

In the following, we illustrate an inference procedure example for a three-order hyperlink, which can be similarly generalized to the high-order hyperlinks following a similar procedure. 
Intuitively, we tend to infer hyperlink status whose underlying probability is either close to 1 or close to 0 due to their 
reduced inference uncertainty. Based on the hyperlink modeling in (\ref{hyper_formula}),
the inference of a hyperlink status follows the three steps: \\
\noindent {\textbf{Step 1:} \textbf{Embedding with observed network:} We first obtain node-wise latent factors based on the observed pairwise links $\Omega_{\bm{Y}}$ and hyperlinks $\Omega_{\mathcal{Y}}$  through
\begin{align}\label{ini_Z}
\bm{Z}^{obs} = \argmin_{\bm{Z}} L(\bm{Z},\Omega_{\bm{Y}},\Omega_{\mathcal{Y}}), 
\end{align} 
where $L(\cdot)$ is the joint embedding loss defined in (\ref{eq3}). The $\bm{Z}^{obs}$ and encodes the position of each node and its geometric relation in the latent space, which is utilized  by the downstream hyperlink inference. The penalty term $\|\bm{Z}\|^2$ is not considered, i.e., $\lambda = 0$ in step 1 for fully utilizing the observed network.
 
\noindent{\textbf{Step 2:} {\textbf{Construct candidate for augmented hyperlink statuses:} Given the prior knowledge such that the probability of establishing a hyperlink among a set of nodes is higher when
there are more observed pairwise links among these nodes compared to those with fewer observed
pairwise links, we construct a candidate pool for augmented hyperlink statuses based on the hierarchical dependence structure and observed pairwise link information. Specifically, we introduce $\Omega_1^{pair} = \{(i,j,k)|Y_{ij} = Y_{ik} = Y_{jk} = 1\}\cap \Omega_{\mathcal{Y}}^c$ and  $\Omega_2^{pair} = \{(i,j,k)|Y_{ij} = Y_{ik} = Y_{jk} = 0\}\cap \Omega_{\mathcal{Y}}^c$ to indicate the candidate pool for augmented hyperlinks and non-hyperlinks, respectively, where $\Omega_{\mathcal{Y}}$ is the observed hyperlink statuses. We denote the total candidate pool as $\Omega_{pool}:=\Omega_1^{pair}\cup \Omega_2^{pair}$.}



\noindent {\textbf{Step 3:} {\textbf{Select informative augmented hyperlink statuses:} We select augmented hyperlink statuses with low inference uncertainty from $\Omega_{pool}$. Based on $\bm{Z}^{obs}$ we estimate the hyperlink probability $\hat{P}_{ijk} = P(Y_{ijk})$ via (\ref{hyper_formula}) for $Y_{ijk}\in \Omega_{pool}$. Note that $\{\hat{P}_{ijk}\}$ only utilizes the information from observed network, which is an independent information source from the hierarchical dependency prior in Step 2. We construct the set of augmented hyperlink statuses $\hat{\Omega}_{\mathcal{Y}}:= \{\hat{Y}_{ijk}\}$ as follows:
\begin{align*}
\hat{Y}_{ijk}  = \begin{cases} 
1, \;\;  (i,j,k)\in \Omega^{pair}_1\;\text{and}\;\hat{P}_{ijk} \geq 1-\delta, \\
0, \;\;  (i,j,k)\in \Omega^{pair}_2\;\text{and}\;\hat{P}_{ijk} \leq \delta,
\end{cases}
\end{align*}  
where $\delta \in (0,0.5)$ is a cut-off that can be tuned on a validation set.   
Finally, we refine the node-wise latent factors $\bm{Z}^{obs}$ through incorporating the set of augmented hyperlink statuses $\Omega_{\mathcal{Y}}\cup \hat{\Omega}_{\mathcal{Y}}$ into the joint embedding loss function in (\ref{eq3}) as $
\bm{Z}^{aug} = \argmin_{\bm{Z}} L(\bm{Z},\Omega_{\bm{Y}},{\Omega_{\mathcal{Y}}\cup \hat{\Omega}_{\mathcal{Y}}})$.}

The effectiveness of the augmentation procedure relies on the existence of dependency between pairwise links and hyperlinks on the
same set of nodes, and the improvement from link augmentation becomes more significant as the degree of dependency increases. Given the dependency exists and it is moderate, we can capture
it via modeling the pairwise links in (\ref{eq1})  and hyperlinks in (\ref{eq2}) on shared latent factors. When the dependency 
is stronger in that two types of links are still correlated conditioning on the latent positions, then
the augmentation procedure can incorporate the conditional dependency via imposing subgroup
structure on the latent positions; and introducing structured regularization on the parameter space
of latent factors can enhance the estimation efficiency.

\subsection{Embeddings Estimation} 

We embed each node into a latent feature $\bm{Z}$ estimated by
minimizing
the joint loss function in \eqref{eq3}. In contrast to existing pairwise link or hyperlink embedding approaches 
involving matrix factorization \citep{menon2011link, qiu2018network,ahmed2013distributed,belkin2002laplacian,cao2015grarep,ou2016asymmetric, zhou2007learning,zhu2016heterogeneous} or tensor decomposition, which could entail high computational cost especially when the order of tensor $m$ is high. Notice that the hyperlink tensor $\mathbf{Y}$ is super-symmetric in that $Y_{i_1i_2\cdots i_m} = Y_{\varphi({i_1})\varphi({i_2})\cdots \varphi({i_m})}$, where $\varphi$ is the order permutation mapping. On this ground, 
we develop a scalable algorithm to minimize \eqref{eq3} while taking advantage of the super-symmetry of a 
hyperlink tensor to reduce the computational cost. 

In general, we estimate the embedding of nodes $\bm{Z}$ through the coordinate gradient descent algorithm where both the gradient and Hessian matrix have explicit forms.
Alternatively, we can replace the gradient update by the first-order gradient descent with an adaptive learning rate strategy such as ADAM \citep{kingma2014adam} to accelerate the convergence and avoid local minima. The detailed estimation procedure is summarized as Gradient Descent Algorithm with Parallel Computing in the supplemental material. One advantage of the proposed algorithm is that it utilizes the  super-symmetry property of a hyperlink tensor to  update the latent vectors corresponding to different tensor modes simultaneously instead of updating each mode iteratively. In addition, updating gradients of the node-wise latent vectors can be performed independently of each other, which makes it feasible for parallelization to accelerate computation. 
\vspace*{-2mm}
\section{Theoretical Results}\label{2.3}
In this section, we investigate the theoretical properties for the proposed joint embedding method. Specifically, we focus on establishing the asymptotic properties 
of link probability estimation as it directly associates with prediction accuracy. 
We consider a class of link-generating processes where the pairwise link statuses $\mathbf{Y} = \{Y_{ij}\}$ and $m$-order hyperlink statuses $\bm{\mathcal{Y}} = \{Y_{i_1i_2\cdot i_m}\}$ can be hierarchically dependent conditioning on the latent position $\bm{Z}$.

Given the node-wise latent factors $\bm{Z} = \{Z_i\}_{i=1}^N$, we decompose the joint distribution as $\mathbf{P}(\bm{\mathcal{Y}},\mathbf{Y}|\bm{Z}) = \mathbf{P}(\bm{\mathcal{Y}}|\mathbf{Y},\bm{Z})\mathbf{P}(\mathbf{Y}|\bm{Z}),$ and assume that the pairwise links $\{\mathbf{P}(Y_{ij}|\bm{Z})\}$ independently follow
\begin{align} \label{link_form_1}
Y_{ij}\sim \text{Bern}\big(\theta_{ij}\big),\;\text{where}\; 
\theta_{ij} = \frac{\exp(Z_iZ_j^T)}{1+\exp(Z_iZ_j^T)};\; 1\leq i<j\leq N.
\end{align} 
For the generation of a $m$-order hyperlink status $\{Y_{i_1i_2\cdots i_m}\}$, we introduce a clique indicator variable $\bigtriangleup_{i_1i_2\cdots i_m}$ as 
\begin{align}\label{clique}
\bigtriangleup_{i_1i_2\cdots i_m} = \begin{cases}
1, \; \text{if}\;Y_{ij} = 1, \; (i,j)\in \{i_1,i_2,\cdots,i_m\},\\
0, \; \text{otherwise}.
\end{cases}
\end{align}
In other words, $\bigtriangleup_{i_1i_2\cdots i_m}$ captures whether all nodes $i_1,i_2,\cdots,i_m$ are pairwisely connected. We assume that $Y_{i_1i_2\cdots i_m}$ generates from a Bernoulli distribution as
\begin{align}\label{link_form_2}
Y_{i_1i_2\cdots i_m}\!\sim\!\mathbf{P}(Y_{i_1i_2\cdots i_m}|\mathbf{Y},\bm{Z})\! = \!\text{Bern}\Big(P(Y_{i_1i_2\cdots i_m}\!=\!1|\bigtriangleup_{i_1i_2\cdots i_m},\bm{Z})\Big),1\!\leq\! i_1\!<\!\cdots\!<\!i_m\!\leq \! N,  
\end{align}
where the probability relies on the clique indicator $\bigtriangleup_{i_1i_2\cdots i_m}$, therefore introduces an additional hierarchical dependency between the pairwise links and hyperlinks in addition to sharing the same set of latent position $\bm{Z}$. 
We provide an explicit formulation of $P(Y_{i_1i_2\cdots i_m}\!=\!1|\bigtriangleup_{i_1i_2\cdots i_m},\bm{Z})$ in the supplementary.
We model the marginal distribution $\mathbf{P}(\bm{\mathcal{Y}}|\bm{Z}) = \{\mathbf{P}(Y_{i_1i_2\cdots i_m}|\bm{Z})\}$ by Bernoulli distribution with probability:
\begin{align}  \label{link_form_3}
\theta_{i_1i_2\cdots i_m}  = P(Y_{i_1i_2\cdots i_m}=1|\bm{Z}) =  \sigma\big[ \sum_{(i,j)\in \{i_1,i_2,\cdots,i_m \}}Z_iZ_j^T + \beta f(Z_{i_1}, Z_{i_2}\cdots Z_{i_m})\big],
\end{align} 
which has the same formulation as (\ref{hyper_formula}) in Section 3.2. Therefore, given pairwise links and hyperlinks follows the joint distribution $\mathbf{P}(\bm{\mathcal{Y}},\mathbf{Y}|\bm{Z})$ as (\ref{link_form_1}) and (\ref{link_form_2}), the \text{JLE} objective function combining (\ref{link_form_1}) and (\ref{link_form_3}) serves as the composite likelihood for $\mathbf{P}(\bm{\mathcal{Y}}|\bm{Z})\mathbf{P}(\mathbf{Y}|\bm{Z})$ as a surrogate of $\mathbf{P}(\bm{\mathcal{Y}},\mathbf{Y}|\bm{Z})$ \citep{lindsay1988composite, varin2011overview}.

{It is natural to consider a non-negative dependency in (\ref{link_form_2}) in that the probability of $Y_{i_1i_2\cdots, i_m}=1$ is higher given that a clique exists among nodes $\{i_1,\cdots,i_m\}$, and the degree of dependency can be quantified as:
\begin{align}\label{link_form_4}
\rho_{i_1i_2\cdots i_m} := P(Y_{i_1i_2\cdots i_m}=1|\bigtriangleup_{i_1i_2\cdots i_m} = 1,\bm{Z})- P(Y_{i_1i_2\cdots i_m}=1|\bigtriangleup_{i_1i_2\cdots i_m} = 0,\bm{Z}), 
\end{align}
where $0\leq \rho_{i_1i_2\cdots i_m} \leq 1$. In practice, a positive dependency between a hyperlink $Y_{i_1i_2\cdots i_m}$ and the corresponding clique $\bigtriangleup_{i_1i_2\cdots i_m}$ in (\ref{link_form_2}) is extensively observed within complex networks. For example, the existence of a protein complex is highly associated with a densely connected subgraph within protein-protein interaction networks \citep{xu2013protein, ma2017identification}. In addition, a high-order social interaction is generally built upon pairwise interactions in social networks \citep{cencetti2021temporal, alvarez2021evolutionary}.}

{In addition, instead of assuming the link statuses are randomly observed within the pairwise link network, we allow that the observation dependency exists and is formulated as:
{\begin{align*}
\rho_{obs}: = corr\Big\{\mathbf{1}_{\{Y_{ij},Y_{ik}\;\text{are observed}\}}(Y_{ij},Y_{ik}), \mathbf{1}_{\{Y_{jk}\;\text{is observed}\}}(Y_{jk})\Big\}.
\end{align*}
Specifically, we consider the case when $\rho_{obs}\geq 0$, and $\rho_{obs}$ measures an increase in the probability of $Y_{jk}$ being observed compared to being randomly observed given $Y_{ij}$ and $Y_{ik}$ are observed.} In many applications, the observation of networks is missing not at random in that the probability of a link's missingness is related to its neighbourhood links connecting to the same nodes \citep{huisman2009imputation}. Therefore, $\rho_{obs}$ quantifies the missing pattern information from the pairwise links.}   

Given the parameter set $\bm{\Theta} = \{\theta_{ij}, \theta_{i_1i_2\cdots i_m};\;  1\leq i< j\leq N,\; 1\leq i_1<\cdots < i_m \leq N \}\in \mathcal{S}\subseteq R^{N\times N} \cup R^{N^m}$, the link prediction accuracy of the proposed method can be established through investigating the convergence property of the $\bm{\Theta}$ estimator. In the following, we establish the consistency and convergence rate of the $\bm{\Theta}$ estimator that minimizes the proposed joint embedding loss function:
\begin{align}\label{formula_11}
l_{joint}(\bm{\Theta}(\bm{Z}); \bm{Y}, 
\mathcal{Y}) = l_{pair}(\bm{\Theta}(\bm{Z}); \bm{Y}) + l_{hyper}(\bm{\Theta}(\bm{Z}); \mathcal{Y})  + \lambda \|\bm{Z}\|^2,  
\end{align}
where $l_{pair}(\bm{\Theta}; \bm{Y})$ and $l_{hyper}(\bm{\Theta}; \mathcal{Y})$ are the loss functions in (\ref{eq1}) and (\ref{eq2}) representing the pairwise link and hyperlink embedding, respectively. Because of the non-convex nature of the loss function (\ref{formula_11}) with respect to $\bm{\Theta}$, obtaining the global minimizer of (\ref{formula_11}) is generally infeasible or computationally expensive in practice. Instead, we establish the convergence property for a class of alternative estimators $\hat{\bm{\Theta}}$ satisfying
\begin{align}\label{formula_2}
l_{joint}(\bm{\hat{\Theta}}; \bm{Y}, 
\mathcal{Y}) \leq \inf_{\Theta \in \mathcal{S}} l_{joint}(\bm{\Theta}; \bm{Y}, 
\mathcal{Y}) + \tau, 
\end{align}
where $\tau$ goes to zero as the number of observed links increases. We show that the alternative estimator $\hat{\bm{\Theta}}$ in (\ref{formula_2}) is still able to achieve the convergence rate of the global minimizer of (\ref{formula_11}) as the sample size increases. We introduce the following regularity assumption: } 


%
%

%

{\noindent (C1): The node-wise latent factors are uniformly bounded such that $\|\bm{Z}\|_{\infty}\leq C$ for some positive constant $C$, where $\|\cdot\|_{\infty}$ indicates the infinity norm. The weighting parameter $\beta$ is bounded such that $1\leq \beta \leq C'$ for some positive constant $C'$.} 

{\noindent\textbf{Remark 4.1}: Assumption (C1) requires that the underlying search space for the latent factors is bounded, hence the parameter space $\mathcal{S}$ is also uniformly bounded. This is a standard assumption for latent factor modeling.}

{In the following, we establish the consistency and convergence rates for the estimator $\bm{\hat{\Theta}}$ from the proposed joint link embedding method (\text{JLE}) when the pairwise links and hyperlinks are hierarchically dependent on each other given their latent information. We first introduce the following quantity: 
\begin{align}\label{C_0}
C_0 := \max_{\{(i,j)|Y_{ij}\in \Omega_{\mathbf{Y}}\}}\Big\{\sum_{\{(i_1,i_2\cdots,i_m):Y_{i_1i_2\cdots i_m}\in \Omega_{\mathcal{Y}}\}}\mathbf{1}_{\{i_1,i_2,\cdots, i_m\}}\big(\{i,j\}\big)\Big\},
\end{align}
where $\mathbf{1}_{A}(e) = 1$ if $e\subseteq A$ and $\mathbf{1}_{A}(e) = 0$ otherwise. Intuitively, $C_0$ measures the largest number of observed hyperlinks such that these hyperlinks overlap in the same pairwise link, and therefore measures the degree of overlapping between the observed pairwise network and the hyperlink network. Notice that $0\leq C_0\leq \min\{\binom{N-2}{m-2},|\Omega_{\mathcal{Y}}|\}$ as $C_0 \asymp \frac{|\Omega_{\mathcal{Y}}|m(m-1)}{N(N-1)}$, given that the observed hyperlink statuses are sampled uniformly from all 
the $C_{N}^m$ possible tuples of nodes, where $\asymp$ denotes the same order. Without loss of generality , we assume that $\{\rho_{i_1i_2\cdots i_m}\}$ are all equal and denoted as $\rho$. Denote $n_{N,m}: = {C}_{N}^2+{C}_{N}^m = \frac{N(N-1)}{2} + \frac{N!}{m!(N-m)!}$. In the following, we establish the consistency and convergence rate for the \text{JLE} using the observed network.}

\begin{theorem} \label{theorem1}
Denote $\bm{\Theta}_0$ as the underlying true link probabilities, and $n_{N,m}: = {C}_{N}^2+{C}_{N}^m$. Under the assumption (C1) and $\tau = O(\epsilon^2)$, we establish the convergence rate for a \text{JLE} estimator $\hat{\bm{\Theta}}$. That is, 
$$
P\left( \frac{\|\hat{\bm{\Theta}} -  \bm{\Theta}_0\|_F}{\sqrt{n_{N,m}}} \geq \eta\right) \leq 11 \exp \left(-c\frac{|\Omega_{\bm{Y}}|+|\Omega_{\mathcal{Y}}|}{(1+C_0\rho)^2}\eta^{2}\right),
$$
where $\|\cdot\|_{F}$ indicates the Frobenius norm, $c \geq 0$ is a constant, $\eta =\max \left(\varepsilon, \lambda^{1 / 2}\right)$, $C_0$ is the degree of overlap links in (\ref{C_0}), and the best possible rate $\varepsilon\sim {\left(\frac{1+C_0\rho}{(|\Omega_{\bm{Y}}|+|\Omega_{\mathcal{Y}}|)^{1 / 2}}\right)}$ is achieved when $\lambda \asymp \varepsilon^{2}$.
\end{theorem}
The Theorem 4.1 provides the convergence property for the \text{JLE} under a broad class of link-generating models which include both scenarios when the hyperlink statuses and pairwise link statuses are independent or hierarchically dependent given the latent position $\bm{Z}$. Specifically, Theorem \ref{theorem1} states that given the magnitude of penalty term $\|\bm{Z}\|$ in (\ref{formula_11}) shrinking to zero with an appropriate rate, the proposed estimator for $\bm{\Theta}$ can achieve the convergence rate at $\frac{1+C_0\rho}{(|\Omega_{\bm{Y}}|+|\Omega_{\mathcal{Y}}|)^{1 / 2}}$. {In addition, Theorem 4.1 shows that the \text{JLE} estimator achieves a faster convergence rate compared to methods using either pairwise link statuses $\Omega_{\mathbf{Y}}$ only (\text{PLE}), or hyperlink statuses $\Omega_{\mathcal{Y}}$ only (\text{HLE}) with the corresponding rate of $\frac{1}{|\Omega_{\mathbf{Y}}|^{1 / 2}}$ or $\frac{1}{|\Omega_{\mathcal{Y}}|^{1 / 2}}$, when the observed pairwise and hyperlinks are conditionally independent, i.e., $\rho=0$ or $C_0 = 0$. When these two types of links are hierarchically dependent, i.e., $\rho>0$, the convergence rates of \text{PLE} and \text{HLE} are $\frac{1}{|\Omega_{\mathbf{Y}}|^{1/2}}$ and $\frac{1+C_0\rho}{|\Omega_{\mathcal{Y}}|^{1/2}}$, respectively.} Therefore, the \text{JLE} estimator still converges faster than \text{HLE} and \text{PLE} when $\frac{|\Omega_{\mathcal{Y}}|}{|\Omega_{\mathbf{Y}}|}\geq 2C_0\rho + C_0^2\rho^2$. We provide discussion on the comparison between existing convergence rate of low-rank matrix completion and Theorem 4.1 in Appendix.

{In the following Theorem 4.2, we further develop the convergence property of the proposed joint embedding estimator incorporating augmented hyperlink statuses. 
{We first define the candidate pool of augmented hyperlink statuses $\Omega_{\text{clique}}\cup \Omega_{\text{non-clique}}$ as
\begin{align}\label{clique_1}
\Omega_{\text{clique}}:= \big\{ Y_{i_1i_2\cdots i_m}|Y_{ij}=1,\{i,j\}\subset\{i_1,i_2,\cdots,i_m\} \big\}\cap \Omega_{pool}, \nonumber \\
\Omega_{\text{non-clique}}:= \big\{ Y_{i_1i_2\cdots i_m}|Y_{ij}=0,\{i,j\}\subset\{i_1,i_2,\cdots,i_m\} \big\}\cap \Omega_{pool}, 
\end{align}
where $\Omega_{pool}$ representing the set $\big\{Y_{i_1i_2\cdots i_m}|Y_{ij}\;\text{is observed}, \{i,j\}\subset \{i_1,i_2,\cdots,i_m\}\big\}$. In other words, we infer the potential hyperlink statuses on the node tuple $\{i_1,i_2,\cdots,i_m\}$, where the pairwise link statuses among $\{i_1,i_2,\cdots,i_m\}$ are observed, and are either all linked together or not linked at all. Given the set of observed pairwise link statuses $\Omega_{\mathbf{Y}}$, the size of $\Omega_{pool}$ is determined by the observation dependency $\rho_{obs}$ among pairwise link statuses as 
\begin{align}\label{clique_2}
|\Omega_{pool}(\rho_{obs})|= O\Big(\frac{|\Omega_{\mathbf{Y}}|^{m-1}}{N^{m-2}(m-1)!}\{\kappa + \rho_{obs}(1-\kappa)\}^{ \binom{m-1}{2}}\Big),
\end{align}
where $\kappa = \frac{2|\Omega_{\mathbf{Y}}|}{N^2-N}\in (0,1)$ is the observation density in the pairwise link network. To ensure $|\Omega_{pool}(\rho_{obs})|>1$, we consider the case where the order of hyperlink is bounded in that $m=O(\log N/\log \kappa^{-1})$. Therefore, 
the size of the candidate pool of hyperlink statuses for augmentation $|\Omega_{\text{clique}}\cup \Omega_{\text{non-clique}}|$ increases as $\rho_{obs}$ becomes larger due to more cliques $\big\{ Y_{i_1i_2\cdots i_m}|Y_{ij}=1,\{i,j\}\subset\{i_1,i_2,\cdots,i_m\} \big\}$ or non-cliques $\big\{ Y_{i_1i_2\cdots i_m}|Y_{ij}=0,\{i,j\}\subset\{i_1,i_2,\cdots,i_m\} \big\}$ being observed with a larger $\rho_{obs}$.} 

To quantify the number of augmented hyperlink statuses within the candidate pool, we introduce two monotone increasing distribution functions $f_{\text{traid}}(\varphi),\;f_{\text{non-traid}}(\varphi): [0,1]\rightarrow [0,1]$ to indicate the proportions:
\begin{align}\label{prop_fun}
f_{\text{clique}}(\varphi) = \frac{|\{Y_{i_1i_2\cdots i_m}\in \Omega_{\text{clique}}| P(Y_{i_1i_2\cdots i_m}=1|\bm{Z})\in [1-\varphi,1)\}|}{| \Omega_{\text{clique}}|},\nonumber \\
f_{\text{non-clique}}(\varphi) = \frac{|\{Y_{i_1i_2\cdots i_m}\in \Omega_{\text{non-clique}}| P(Y_{i_1i_2\cdots i_m}=1|\bm{Z})\in (0,\varphi]\}|}{|\Omega_{\text{non-clique}}|}.
\end{align}
Intuitively, $f_{\text{clique}}(\cdot)$ and $f_{\text{non-clique}}(\cdot)$ quantify the proportion of hyperlinks with a large generating probability within $\Omega_{\text{clique}}$ and hyperlinks with small generating probability within $\Omega_{\text{non-clique}}$. Based on (\ref{clique_1}), (\ref{clique_2}) and (\ref{prop_fun}), we can show that the size of augmented hyperlink statuses with inference bias smaller than $\epsilon$ is formulated as:
\begin{align*}
|\hat{\Omega}_{\mathcal{Y}}(\epsilon,\rho_{obs},\rho)| = f_{\text{clique}}\big(\min\{\frac{\epsilon}{1-\rho},1\}\big)|\Omega_{\text{clique}}(\rho_{obs})| + f_{\text{non-clique}}\big(\min\{\frac{\epsilon}{1-\rho},1\}\big)|\Omega_{\text{non-clique}}(\rho_{obs})|.
\end{align*}       
Let $\epsilon_{ini}$ denote the mean square error of hyperlink probability estimation from the \text{JLE} using observed network, the convergence property for the estimator of joint link embedding incorporating augmented hyperlink statuses is established in the following Theorem 4.2.

{\begin{theorem}\label{theorem2}
Assume that the hyperlink statuses are inferred through the proposed augmentation procedure in Section 3.4. Denote $\hat{\bm{\Theta}}$ as the minimizer of the JLE objective function with augmented hyperlink statuses, and $n_{N,m}: = {C}_{N}^2+{C}_{N}^m$. Under the assumption (C1) and $\tau = O(\epsilon^2_{aug})$, we establish the convergence rate for $\hat{\bm{\Theta}}$: 
\begin{align}\label{4.2}
P \left(\! \frac{\|\hat{\bm{\Theta}} -  \bm{\Theta}_{0}\|_{F}}{\sqrt{n_{N,m}}}
\geq \!\eta\! \!\right)\!\! \leq \! 11 \exp \!\left\{
\!-c\frac{|\Omega_{\mathbf{Y}}|+\!|\Omega_{\mathcal{Y}}| \!+\! |\hat{\Omega}_{\mathcal{Y}}(\epsilon_{aug},\rho_{obs},\rho)|}{(1+C_0\rho)^2}\Big(\!1
\!+\! c_1 \!+\! c_1\frac{\epsilon_{ini}^2}{\epsilon_{aug}^2}\!\Big)^{-1}\!\!\! \eta^{2} \right\},
\end{align}
where $c$ is the same constant as in Theorem 4.1, $\eta =\max \left(\varepsilon_{aug},\lambda^{1 / 2}\right)$, $C_0$ is the degree of link overlap in (\ref{C_0}), and 
$c_1 = \frac{\max\{|\hat{\Omega}_{\mathcal{Y}}((1-\rho)\epsilon_{ini},\rho_{obs},\rho)|-|\hat{\Omega}_{\mathcal{Y}}(\epsilon_{aug},\rho_{obs},\rho)| ,0\}}{|\Omega_{\mathbf{Y}}|+ |\Omega_{\mathcal{Y}}| + |\hat{\Omega}_{\mathcal{Y}}(\epsilon_{aug},\rho_{obs},\rho)|}$.
The best possible rate can be achieved when $\lambda \asymp \varepsilon_{aug}^2$ with $\epsilon_{aug}\sim \frac{1+C_0\rho}{(|\Omega_{\mathbf{Y}}|+ |\Omega_{\mathcal{Y}}| + |\hat{\Omega}_{\mathcal{Y}}(\epsilon_{aug},\rho_{obs},\rho)|)^{1/2}}$.
\end{theorem}
{Theorem 4.2 shows that the proposed \text{JLE} can achieve a faster convergence rate via incorporating the augmented hyperlink statuses. Compared with the convergence rate of \text{JLE} at $\frac{1+C_0\rho}{(|\Omega_{\bm{Y}}|+|\Omega_{\mathcal{Y}}|)^{1 / 2}}$ from Theorem 4.1, the augmented $\text{JLE}$ is $\sqrt{1 + \frac{|\hat{\Omega}_{\mathcal{Y}}(\epsilon_{aug},\rho_{obs},\rho)|}{|\Omega_{\bm{Y}}|+|\Omega_{\mathcal{Y}}|}}$ times faster, and the improvement increases as the link dependency $\rho$ and observation dependency $\rho_{obs}$ increase.}  
{In addition, when the estimation bias $\epsilon_{aug}$ is larger than $(1-\rho)\epsilon_{ini}$, the term $(1+c_1 + c_1\frac{\epsilon^2_{ini}}{\epsilon^2_{aug}})^{-1}$ degenerates to $1$. For an $\epsilon_{aug}$ smaller than $(1-\rho)\epsilon_{ini}$, the term $(1+c_1 + c_1\frac{\epsilon^2_{ini}}{\epsilon^2_{aug}})^{-1}$ decreases as 
the $\epsilon_{aug}$ becomes smaller, hence reducing the probability of achieving the convergence rate in (\ref{4.2}). Intuitively, the quantity 
$\epsilon_{ini}/\epsilon_{aug}$ reflects the trade-off between bias and variance in that the hyperlink augmentation introduces both additional signals and noise in selecting an informative augmented hyperlink from $\Omega_{pool}$. However, Theorem 4.2 implies that the augmented \text{JLE} can achieve the best convergence rate with a higher probability than \text{JLE} as long as $\Big\{\big( f_{\text{clique}}(\epsilon_{ini}) - f_{\text{clique}}(\frac{\epsilon_{aug}}{1-\rho}) \big) + \big(f_{\text{non-clique}}(\epsilon_{ini}) - f_{\text{non-clique}}(\frac{\epsilon_{aug}}{1-\rho})\big) \Big\}\frac{\epsilon_{ini}^2}{\epsilon_{aug}^2} = O(\frac{|\Omega_{\mathbf{Y}}|+|\Omega_{\mathcal{Y}}| + |\hat{\Omega}_{\mathcal{Y}}(\epsilon_{aug},\rho_{obs},\rho)|}{|\Omega_{\text{clique}}|+|\Omega_{\text{non-clique}}|})$, which can be satisfied in general when $\rho$ or the size of observed links $|\Omega_{\mathbf{Y}}|+|\Omega_{\mathcal{Y}}|$ is larger.} Asymptotically, the \text{JLE} estimator and the
augmented \text{JLE} estimator achieve the same convergence rate towards the true link probability. However, the augmented \text{JLE} is more efficient than the regular \text{JLE}.

In addition, Theorem 4.2 implies that the augmented \text{JLE} performs better when the pairwise link status and hyperlink status are conditionally dependent compared to the case when they are conditionally independent. Specifically, the convergence rate of augmented \text{JLE} is $\epsilon_{aug}(0)\sim\frac{1}{(|\Omega_{\mathbf{Y}}|+ |\Omega_{\mathcal{Y}}| + |\hat{\Omega}_{\mathcal{Y}}(\epsilon_{aug},\rho_{obs},0)|)^{1/2}}$ under the independent model and is $\epsilon_{aug}(\rho)\sim\frac{1+C_0\rho}{(|\Omega_{\mathbf{Y}}|+ |\Omega_{\mathcal{Y}}| + |\hat{\Omega}_{\mathcal{Y}}(\epsilon_{aug},\rho_{obs},\rho)|)^{1/2}}$ under dependent model. We can show that $\epsilon_{aug}(\rho)< \epsilon_{aug}(0)$ given that 
\begin{align*}
\lim_{\varphi \to 0}\frac{f_{\text{clique}}(\varphi)}{\varphi^{2C_0(L+1)}} = 0\;\;\text{and}\;\; \lim_{\varphi \to 0}\frac{f_{\text{non-clique}}(\varphi)}{\varphi^{2C_0(L+1)}} = 0, 
\end{align*} 
i.e., both the $f_{\text{clique}}(\varphi)$ and $f_{\text{non-clique}}(\varphi)$ do not have a heavy left-tail, where $L:=\frac{|\Omega_{\mathbf{Y}}|+ |\Omega_{\mathcal{Y}}|}{|\hat{\Omega}_{\mathcal{Y}}(\epsilon_{aug},\rho_{obs},0)|}$ and $L$ decreases as observation dependency $\rho_{obs}$ increases. Furthermore, the selection bias term $1 + c_1 + c_1\frac{\epsilon_{ini}^2}{\epsilon_{aug}^2}$ in (\ref{4.2}) becomes smaller when $\rho>0$. Therefore, the augmented \text{JLE} achieves a faster convergence rate with a higher probability when the pairwise links and hyperlinks are conditional dependent compared with independent model. We provide discussion on the generalization of Theorem 4.2 in supplementary material.

\color{black}

\color{black}
       
\section{Numerical Study}\label{2.4}

In this section, we conduct simulation studies to illustrate the performance of the proposed method on pairwise link and hyperlink predictions on a network. {In particular, we investigate three scenarios where the incorporated hyperlinks are directly observed, and are augmented through the observed network. In addition, we consider the scenario when hyperlink statuses are dependent on pairwise link statuses conditioning on the latent factors.}

\subsection{Study 1: Link Predictions with Observed Hyperlinks}

In the first simulation study, we consider the network generated from an underlying latent space model to compare the performance of various methods based on different criteria of prediction performance.

Suppose there are $N$ nodes in a network and the rank of nodewise latent factors $\bm{Z} = \{Z_i\}_{i=1}^N$ is 5, where $Z_i = (Z_{i1},Z_{i2},Z_{i3},Z_{i4},Z_{i5})$, and each latent feature $Z_{ir}$ is generated from a mixture uniform distribution $Z_{ir} \sim \mu\times unif(-1,-0.6) + (1-\mu)\times unif(0.6,1),\; i = 1,\cdots,N,\;r = 1,\cdots,5,$
with $\mu \sim Bern(1,0.5)$.
{We introduce a feature weighting vector $\alpha = (1,1,1,0.2,0.2)$ and the weighted latent feature $\bm{Z_{\alpha}} = \{Z_{\alpha,i}\}_{i=1}^N$ where $Z_{\alpha,i} = (\alpha_1 Z_{i1},\alpha_2 Z_{i2}, \alpha_3 Z_{i3},\alpha_4 Z_{i4},\alpha_5 Z_{i5})$.} 
Given the latent factors $\bm{Z_{\alpha}}$, pairwise link statuses is generated from the latent factor model in (\ref{link_form_1}). Once pairwise link network $Y$ is generated,  we randomly split $Y$ into training, validation and testing sets with corresponding proportions of $60\%$, $20\%$ and $20\%$, respectively.


In the following simulations, we consider a network consisting of $3$-order hyperlinks where each hyperlink status $Y_{ijk}$ is independently generated to mimic the case where hyperlinks are directly observed, that is,
{\begin{align*}
Y_{ijk} \sim Bern(P_{ijk}),
P_{ijk} = \sigma\big\{c(Z_{\alpha,i}^TZ_{\alpha,j}+Z_{\alpha,i}^TZ_{\alpha,k}+Z_{\alpha,j}^TZ_{\alpha,k})+\beta f(Z_{\alpha,i},Z_{\alpha,j},Z_{\alpha,k})\big\},
\end{align*} 
where $\alpha = (0.2,0.2,0.2,1,1)$, $c=1$, $\beta=3$}. 
For generating the training and testing sets of hyperlinks, we use a hierarchical sampling procedure and the size of the training $3$-order hyperlinks is about $0.2\%$ of 
all possible hyperlinks. This is consistent with real applications in that the multi-way relations are more difficult to observe and cost more to verify in contrast to the pairwise links. The details about the hierarchical sampling procedure is provided in supplemental material.



We than randomly divide the complement set $\{Y_{ijk}|Y_{ij}\in\Omega_{Y},Y_{ik}\in\Omega_{Y},Y_{jk}\in\Omega_{Y}\}\cap\Omega_{\mathcal{Y}}^c$ into the validation set $\Omega_{\mathcal{Y}}^{valid}$ and the testing set $\Omega_{\mathcal{Y}}^{test}$ with a proportion of $50\%$ and $50\%$. Similar to the pairwise link, we further differentiate the links in the testing set $\Omega_{\mathcal{Y}}^{test}$ through defining
{\small 
\begin{align*}
\mathcal{A}^{hyper}_{1} = \{Y_{ijk} \in \Omega_{\mathcal{Y}}^{test} | 0.2 \leq  P(Y_{ijk}) \leq 0.8\}, 
\mathcal{A}^{hyper}_{2} = \{Y_{ijk} \in \Omega_{\mathcal{Y}}^{test}| P(Y_{ijk}) < 0.2\; \text{or}\;  P(Y_{ijk}) > 0.8\}.
\end{align*}} 
The rationale for such a testing set formulation is to investigate the prediction performance of different methods on links with intrinsic uncertainty at different levels. {We also illustrate performance comparisons on the complete testing sets of pairwise link statuses and hyperlink statuses, which are denoted as $\mathcal{A}^{pair}_{test}$, $\mathcal{A}^{hyper}_{test}$, respectively.}

To investigate the performance of incorporating hyperlinks on improving link prediction, we compare six different methods. The first three methods are based on the proposed framework. Specifically, the first one obtains the estimation of the latent factor $\bm{Z}$ through the loss function in (\ref{eq1}), which is equivalent to only utilizing observed pairwise links for embedding, and is denoted as \textbf{pairwise link embedding (PLE)}. The second method estimates the latent factor $\bm{Z}$ through the loss function defined in (\ref{eq2}) where the latent concordance within the link function is replaced by symmetric CP decomposition $\sum_{r=1}^5Z_{ir}Z_{jr}Z_{kr}$, 
and is denoted as \textbf{hyperlink embedding (HLE)}. The proposed method is denoted as \textbf{joint link embedding (JLE)} which incorporates both pairwise links and hyperlinks to jointly estimate $\bm{Z}$ through the proposed loss function (\ref{eq3}), where we set all the hyperlink weights $\{w_{i_1i_2i_3}\} = 1$. 

In addition, we also compare the proposed method with three other popular and state-of-the-art network embedding methods including learning graph representations with global structural information (GraRep) \citep{cao2015grarep}, large-scale information network embedding (LINE) \citep{tang2015line} and scalable feature learning for networks (Node2Vec) \citep{grover2016node2vec}. In general, they encode observed pairwise or high-order relations into node-wise latent factors through decomposing a series of graphical laplacian matrices with different orders, or  encouraging $\bm{Z}$ to conform with the similarity among nodes calculated based on biased random walks on the observed network. We obtain node-wise embedding estimation through the different methods mentioned above and then predict the probability of testing links following (\ref{pre1}) and (\ref{pre2}). The performance of prediction is measured by the AUC (area under the ROC curve). 


%

\begin{table}[h]
\centering
\caption{{The link prediction performance of different methods on different hyperlink testing set when hyperlinks are observed.}}
\vspace{-3mm}
\resizebox{13cm}{!}{
\begin{tabular}{| c | c | c | c | c | c | c | c | } 
  \hline \hline 
& network size  &  Node2Vec & LINE & GraRep  & PLE & HLE  &  \textbf{JLE}    \\ \hline 
    
%
%
%
%
%
%
%


\multirow{3}{*}{AUC on $\mathcal{A}^{hyper}_1$} & N = 100 & 0.50 &  0.52 & 0.50 &  0.50  & 0.51  &  \textcolor{red}{0.56}  \\
& N = 200 & 0.50 &  0.53 & 0.50 &  0.50 & 0.52  &  \textcolor{red}{0.58}  \\
&N = 300 & 0.50 &  0.53 & 0.50 & 0.50  & 0.51  & \textcolor{red}{0.59} \\ \hline

\multirow{3}{*}{AUC on $\mathcal{A}^{hyper}_2$} & N = 100 & 0.50 &  0.54 & 0.63 & 0.77  & 0.78  & \textcolor{red}{0.92}   \\
 & N = 200 & 0.50 &  0.55 & 0.77 &  0.83 &  0.86 &  \textcolor{red}{0.94}  \\
  & N = 300 & 0.50 &  0.55 & 0.80 & 0.84 & 0.86  & \textcolor{red}{0.95}  \\ \hline
  


%
%
%
%

  \hline
   \hline
\end{tabular}}
\label{tab1}
\end{table}

{The prediction results are provided in Table \ref{tab1} and Figure \ref{fig:5_1}, showing that the proposed joint embedding method consistently outperforms the other methods in terms of achieving higher AUC. 
{For the pairwise link prediction, the proposed joint embedding (\textbf{JLE}) slightly outperforms PLE.} 
For hyperlink prediction, the joint embedding also achieves more than $10\%$ improvement on testing set $\mathcal{A}_1^{hyper}$, and more than $16\%$ improvement on testing set $\mathcal{A}_2^{hyper}$ compared with using hyperlinks (\textbf{HLE}) alone. This indicates that 
the joint embedding strategy takes advantage of dependency to encode both two-way relations and multi-way relations into latent factor $\bm{Z}$ simultaneously, hence achieving better prediction performance, especially for low-uncertainty links.}


\begin{figure}[h]
\centering
   \includegraphics[width=5.6in,height=3in]{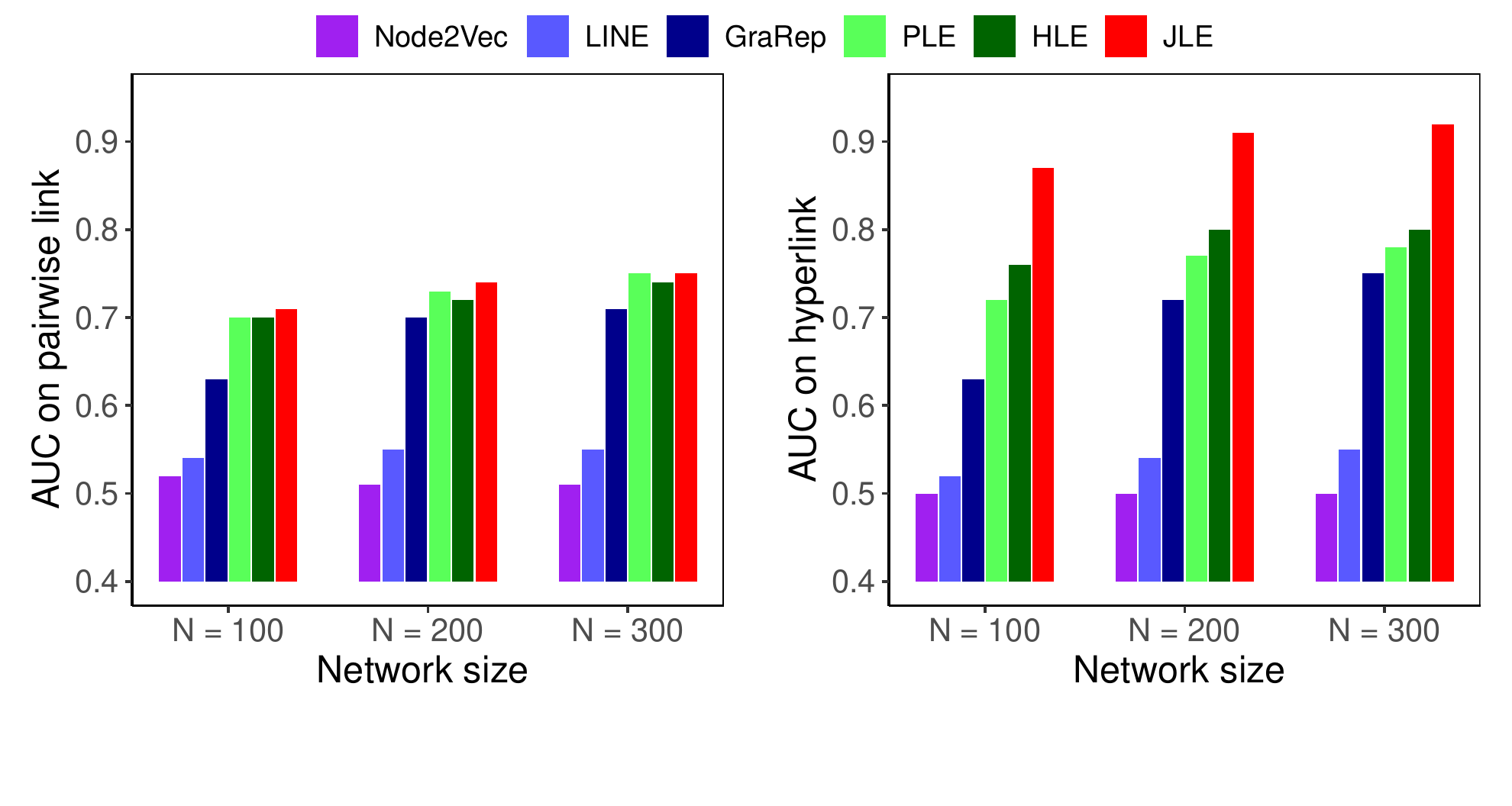}
   \vspace{-10mm}
\caption{{The AUC of link prediction on pairwise link testing set $\mathcal{A}_{test}^{pair}$, and hyperlink testing set $\mathcal{A}_{test}^{hyper}$ when hyperlinks are observed.}}
\label{fig:5_1}
\end{figure} 

Compared with the three other 
existing embedding methods, the proposed method achieves comparable performance to {the best competing method GraRep on $\mathcal{A}_{test}^{pair}$, but more than $12\%$ improvement when $N=100$. In addition, the improvement on hyperlink prediction is more than $16\%$ on $\mathcal{A}_1^{hyper}$ and $\mathcal{A}_2^{hyper}$, and more than $22\%$ on $\mathcal{A}_{test}^{hyper}$.} The improvement with the proposed method demonstrates that the high-order approximation captured by hyperlinks is crucial for inferring potential pairwise links. In terms of hyperlink prediction, the collection of two-way relations alone is not sufficient to determine the underlying multi-way relations. 
{In addition, 
we provide the more detailed simulation results, and the ROC for comparisons between PLE, HLE, and JLE in the supplementary material. }

\subsection{Study 2: Link Prediction with the Hyperlinks Inferred}
In this subsection, we mimic the situation where few hyperlinks are observed, but most of them are inferred from the observed pairwise links. The inferred hyperlinks might be misspecified due to the error propagation from misspecified pairwise links, 
random sampling, and gaps between two-way relations and multi-way relations. 
Nevertheless, the intrinsic hierarchical dependency between hyperlinks and pairwise links can still benefit the recovery of partial high-order information via incorporating information inferred beyond the observed two-way relations. In this study we investigate the case where the pairwise link statuses and hyperlink statuses are independent conditioning on the latent factors $\bm{Z}$.

The training dataset and testing dataset are generated via the same procedure in Study 1. Specifically, the hyperlink training set $\Omega_{\mathcal{Y}}$ consists of $1\%$ random samples from the set $\Omega^{hyper}_{train} = \{Y_{ijk}|Y_{ij}\in\Omega_{Y},Y_{ik}\in\Omega_{Y},Y_{jk}\in\Omega_{Y}\}$, which leads to a sparser hyperlink scenario compared with Study 1. We denote the joint embedding method with augmented hyperlinks proposed in Section 3.4 as \textbf{Aug JLE}. Table \ref{tab2} provides the link prediction performance.

\begin{table}[h]
\centering
\caption{{The link predictions performance of different methods. The performance of prediction is measured by the AUC.}}

\begin{adjustbox}{width=0.9\textwidth}
\begin{tabular}{| c | c | c | c | c | c | c | c | c |  } 
  \hline \hline  
& network size  &  Node2Vec & LINE & GraRep  & PLE  & HLE & JLE &\textbf{Aug JLE}    \\ \hline 
    
%
%
%
%
%
%
%

\multirow{2}{*}{AUC on $\mathcal{A}^{pair}_{test}$} & N = 100 & 0.52 & 0.54  & 0.64 & 0.68 & 0.62 &  0.68  &  \textcolor{red}{0.72}  \\
& N = 200 & 0.51 & 0.55  & 0.71  &  0.75 &  0.69  &  0.74& \textcolor{red}{0.76}\\
\hline

\multirow{2}{*}{AUC on $\mathcal{A}^{hyper}_{test}$} & N = 100 & 0.50 &  0.52 & 0.66  &  0.73 & 0.63 & 0.75  & \textcolor{red}{0.79}    \\
 & N = 200 & 0.50 &  0.54 & 0.78   & 0.83  & 0.74  & 0.81  & \textcolor{red}{0.85} \\
  
%
%

  \hline
   \hline
\end{tabular}
\end{adjustbox}
\label{tab2}
\end{table}

Table \ref{tab2} and Figure \ref{fig:5_2} show that the proposed joint embedding method with augmented hyperlinks achieves the best performance on link prediction compared with embedding methods using observed links only. {In particular, the augmented JLE outperforms PLE and JLE on both $\mathcal{A}_{test}^{pair}$ and $\mathcal{A}_{test}^{hyper}$, and achieves more than $7\%$ improvement on $\mathcal{A}_{test}^{pair}$, and $10\%$ improvement on $\mathcal{A}_{test}^{hyper}$ compared with the other competing methods. 
The simulation results show the advantage of introducing dependency between pairwise link and hyperlink, which is the key step for hyperlink inference. However, the dependency are relatively weak when pairwise links and hyperlinks are independent conditioning on latent factors, which leads to a limited improvement from the hyperlink augmentation procedure. Therefore, we investigate the performance of the proposed methods when the dependency between pairwise links and hyperlinks are stronger. {The detailed simulation results, and the ROC for comparisons between PLE, HLE, JLE, and Aug JLE are included in supplementary material.}

\begin{figure}[h]
\centering
   \includegraphics[width=5.6in,height=2.9in]{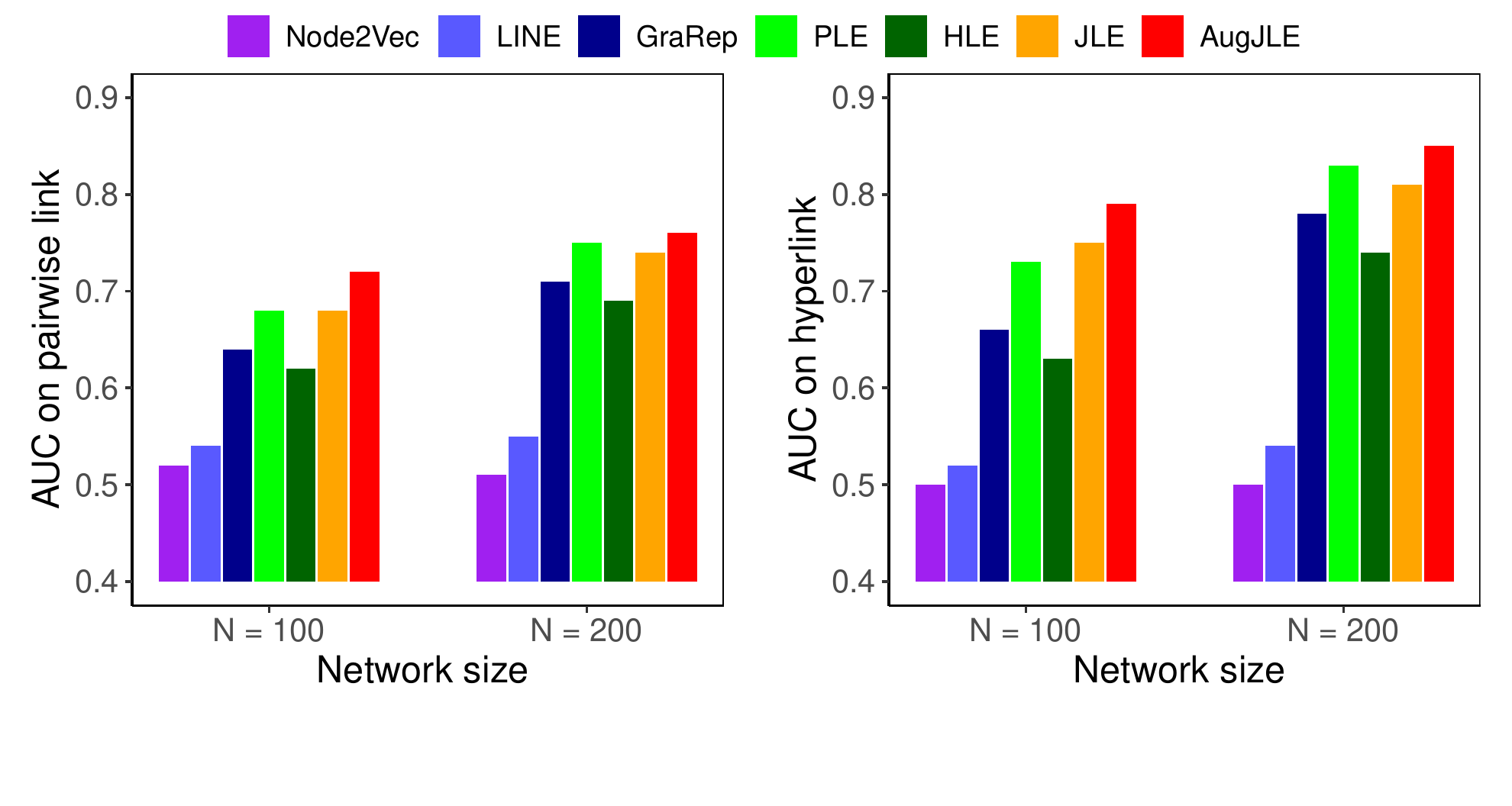}
   \vspace{-10mm}
\caption{{The AUC of link prediction on entire pairwise link testing set $\mathcal{A}_{test}^{pair}$, and entire hyperlink testing set $\mathcal{A}_{test}^{hyper}$ when pairwise links and hyperlinks are conditional independent.}}
\label{fig:5_2}
\end{figure}

{We also investigate the improvement of augmented JLE over JLE in Table \ref{tab3} under different missing rates of pairwise link status and a fixed sample size of hyperlink statuses. The missing rates for the observed pairwise link statuses are set as $20\%$, $30\%$, $40\%$, and $50\%$, and the sample size of hyperlink statuses are $1\%$ of $\Omega^{hyper}_{train}$. 
Table \ref{tab3} shows that the augmented JLE outperforms the JLE
for both link prediction and link probability recovery under various sample sizes, and the improvement from the augmented JLE decreases as the size of observed link status increases.} 

\begin{table}[h]
\centering
\caption{The performance comparison between the augmented JLE and JLE under different missing rates of link status.}
\vspace{-2mm}

\begin{adjustbox}{width=0.9\textwidth}
\begin{tabular}{| c | c | c | c | c | c | c | c | c |  } 
  \hline \hline  
missing rate & \multicolumn{2}{|c|}{$50\%$}  &  \multicolumn{2}{|c|}{$40\%$} & \multicolumn{2}{|c|}{$30\%$} & \multicolumn{2}{|c|}{$20\%$}  
\\ \hline 

& JLE & \textbf{Aug JLE} & JLE & \textbf{Aug JLE} & JLE & \textbf{Aug JLE} & JLE & \textbf{Aug JLE} \\ \hline 
    
%
%
%
%
%
%
%
%
%
%

{AUC on $\mathcal{A}^{pair}_{test}$} & 0.60 & \textcolor{red}{0.65} &  0.64 & \textcolor{red}{0.69} & 0.68 & \textcolor{red}{0.71} &  0.71  &  \textcolor{red}{0.73}  \\ \hline

{AUC on $\mathcal{A}^{hyper}_{test}$} & 0.63 & \textcolor{red}{0.71} & 0.69  & \textcolor{red}{0.76}  &  0.74 & \textcolor{red}{0.79} & 0.78  & \textcolor{red}{0.82}    \\  

%
%
%
%

  \hline
   \hline
\end{tabular}
\end{adjustbox}
\label{tab3}

\end{table}




%
\color{black}

\subsection{Study 3: Link Prediction under the Conditional Dependent Model}

{In the previous simulation studies, the link-generating process follows a conditional independent model in that the pairwise link statuses and hyperlink statuses are independent conditioning on the nodewise latent factors $\bm{Z}$; i.e., we have $\rho_{i_1i_2\cdots i_m} = 0 $ in (\ref{link_form_4}) and $P(Y_{i_1i_2\cdots i_m}=1|\bigtriangleup_{i_1i_2\cdots i_m},\bm{Z}) = P(Y_{i_1i_2\cdots i_m}=1|\bm{Z})$ for $1\leq i_1\leq \cdot i_m \leq N$. In this subsection, we consider a setting where pairwise link statuses and hyperlink statuses are conditional dependent given latent factors $\bm{Z}$, i.e., $\rho_{i_1i_2\cdots i_m} > 0$ and the hyperlink statuses are generated based on both $\bm{Z}$ and the pairwise link clique $\bigtriangleup_{i_1i_2\cdots i_m}$ in (\ref{clique}). We investigate the performance comparisons between the augmented joint link-embedding method (Aug \text{JLE}) and other link-embedding methods using observed networks only.}  

{In the following, we assume that the $N$ latent positions $\bm{Z}\in R^{N\times 5} = \{Z_i\}_{i=1}^N$ evenly distribute across $K=6$ clusters.   
Then the pairwise link network is generated from (\ref{link_form_1}), and we adjust the observation dependency within a pairwise link network so that $\rho_{obs}$ can be equal to different values.
To incorporate the conditional dependency between pairwise link statuses and hyperlink statuses given latent factors, we adopt a random effect model for generating hyperlink statuses. Specifically, we adjust the magnitude of random effect such that empirical estimation of condition dependency (\ref{link_form_4}) is equal to a specific value. We provide the more details regarding the link generating process and settings of dependency in the supplemental material.    

We randomly split pairwise link statuses into training, validation and testing sets with corresponding proportions of $40\%$, $20\%$ and $40\%$, respectively. We consider a sparse hyperlink network scenario in that the hyperlink training set consists of only 30
hyperlink statuses. The hyperlink testing set $\mathcal{A}_{test}^{hyper} = \{Y_{i_1i_2i_3}\}$ are collected in a way that
$\{Y_{i_1i_2i_3}|Y_{i_1i_2}, Y_{i_1i_3}, Y_{i_2i_3}\;\text{are not all observed}\}$, and the hyperlink statuses $|\{Y_{i_1i_2i_3}=1\}|$ and $|\{Y_{i_1i_2i_3}=0\}|$ are balanced.} 

{We increase the link dependency $\rho$ from 0.25 to 0.85, and increase the observation dependency $\rho_{obs}$ from 0.15 to 0.35. We compare the augmented \text{JLE} with other methods on both pairwise link and hyperlink predictions under various degree of dependency. Table \ref{tab_1}, Table \ref{tab_2} and Figure \ref{fig_3} show that the augmented \text{JLE} consistently outperforms the \text{JLE} and other methods for both pairwise link prediction and hyperlink prediction under different sizes of networks. In addition, the improvement from the augmented \text{JLE} is more significant when $\rho$ and $\rho_{obs}$ are large. Specifically, the improvement is up to $33\%$ on pairwise link prediction and $25\%$ on hyperlink prediction over \text{JLE} when $\rho_{obs} = 0.35$ and $\rho = 0.85$. 
Furthermore, Table \ref{tab_3} shows that the link dependency $\rho$ can also affect the performance of the augmented \text{JLE}. That is, given $\rho_{obs}$ is fixed, the improvement from the augmented \text{JLE} is more significant when $\rho$ is large. Intuitively, as the conditional dependency
becomes stronger, the augmentation procedure imposes stronger structural constraints on the latent factor $\bm{Z}$ to capture the underlying subgroup structure on the latent space.}

\begin{table}[h]
\centering
\caption{The link predictions performance of different methods. The performance is
measured by the area under the curve (AUC) and higher is better. The network size is $N = 120$.}

\begin{adjustbox}{width=0.71\textwidth}
\begin{tabular}{| c | c | c | c | c |  } 
  \hline \hline  
  
 $\rho_{obs}= 0.15,\; \rho = 0.25$ & PLE & HLE  & JLE &\textbf{Aug JLE}    \\ \hline

{AUC$_{pair}$} & 0.56(0.02) &  0.52(0.02) & 0.57(0.01)  &  \textcolor{red}{0.61(0.03)}  \\

{AUC$_{hyper}$} &  0.57(0.02) & 0.53(0.01) & 0.57(0.02)  & \textcolor{red}{0.62(0.02)}  \\

\hline

  $\rho_{obs}= 0.25,\; \rho = 0.55$ & PLE & HLE & JLE &\textbf{Aug JLE}    \\ \hline

{AUC$_{pair}$} & 0.55(0.01) & 0.53 (0.02) & 0.59(0.02)  &  \textcolor{red}{0.67(0.02)}  \\

{AUC$_{hyper}$} &  0.59(0.01) & 0.55(0.02) & 0.60(0.02)  & \textcolor{red}{0.68(0.01)}  \\

\hline

$\rho_{obs}= 0.35,\; \rho = 0.85$ & PLE  & HLE & JLE &\textbf{Aug JLE}    \\ \hline

{AUC$_{pair}$} & 0.54(0.02) &  0.51(0.02) & 0.57(0.02)  &  \textcolor{red}{0.68(0.02)}  \\

{AUC$_{hyper}$} &  0.54(0.01) & 0.50(0.03)  & 0.57(0.01)  & \textcolor{red}{0.68(0.01)}  \\
 
\hline

  \hline
   \hline
\end{tabular}
\end{adjustbox}
\label{tab_1}
\end{table}

\begin{table}[h]
\centering
\caption{The link predictions performance of different methods. The performance is
measured by the area under the curve (AUC) and higher is better. The network size is $N = 240$.}

\begin{adjustbox}{width=0.71\textwidth}
\begin{tabular}{| c | c | c | c | c |  } 
  \hline \hline

 $\rho_{obs}= 0.15,\; \rho = 0.25$ & PLE & HLE  & JLE &\textbf{Aug JLE}    \\ \hline

{AUC$_{pair}$} & 0.59(0.01) &  0.52(0.01) & 0.60(0.01)  &  \textcolor{red}{0.67(0.01)}  \\

{AUC$_{hyper}$} &  0.60(0.02) & 0.51(0.01) & 0.60(0.01)  & \textcolor{red}{0.65(0.01)}  \\
 \hline

  $\rho_{obs}= 0.25,\; \rho = 0.55$ & PLE & HLE & JLE &\textbf{Aug JLE}    \\ \hline

{AUC$_{pair}$} & 0.59(0.01) & 0.56 (0.01) & 0.60(0.01)  &  \textcolor{red}{0.71(0.01)}  \\

{AUC$_{hyper}$} &  0.63(0.02) & 0.53(0.01) & 0.62(0.01)  & \textcolor{red}{0.71(0.01)}  \\
\hline

$\rho_{obs}= 0.35,\; \rho = 0.85$ & PLE  & HLE & JLE &\textbf{Aug JLE}    \\ \hline

{AUC$_{pair}$} & 0.54(0.01) &  0.52(0.01) & 0.54(0.01)  &  \textcolor{red}{0.72(0.01)}  \\

{AUC$_{hyper}$} &  0.54(0.02) & 0.50(0.01)  & 0.56(0.01)  & \textcolor{red}{0.70(0.01)}  \\

\hline

  \hline
   \hline
\end{tabular}
\end{adjustbox}
\label{tab_2}
\end{table}

\begin{figure}[h]
\caption{{\small The link predictions performance of different link embedding methods. The network size is $N = 240$.}}
\vspace{-6mm} 
       				\begin{center}
      					\includegraphics[width=5.9in,height=2.3in]{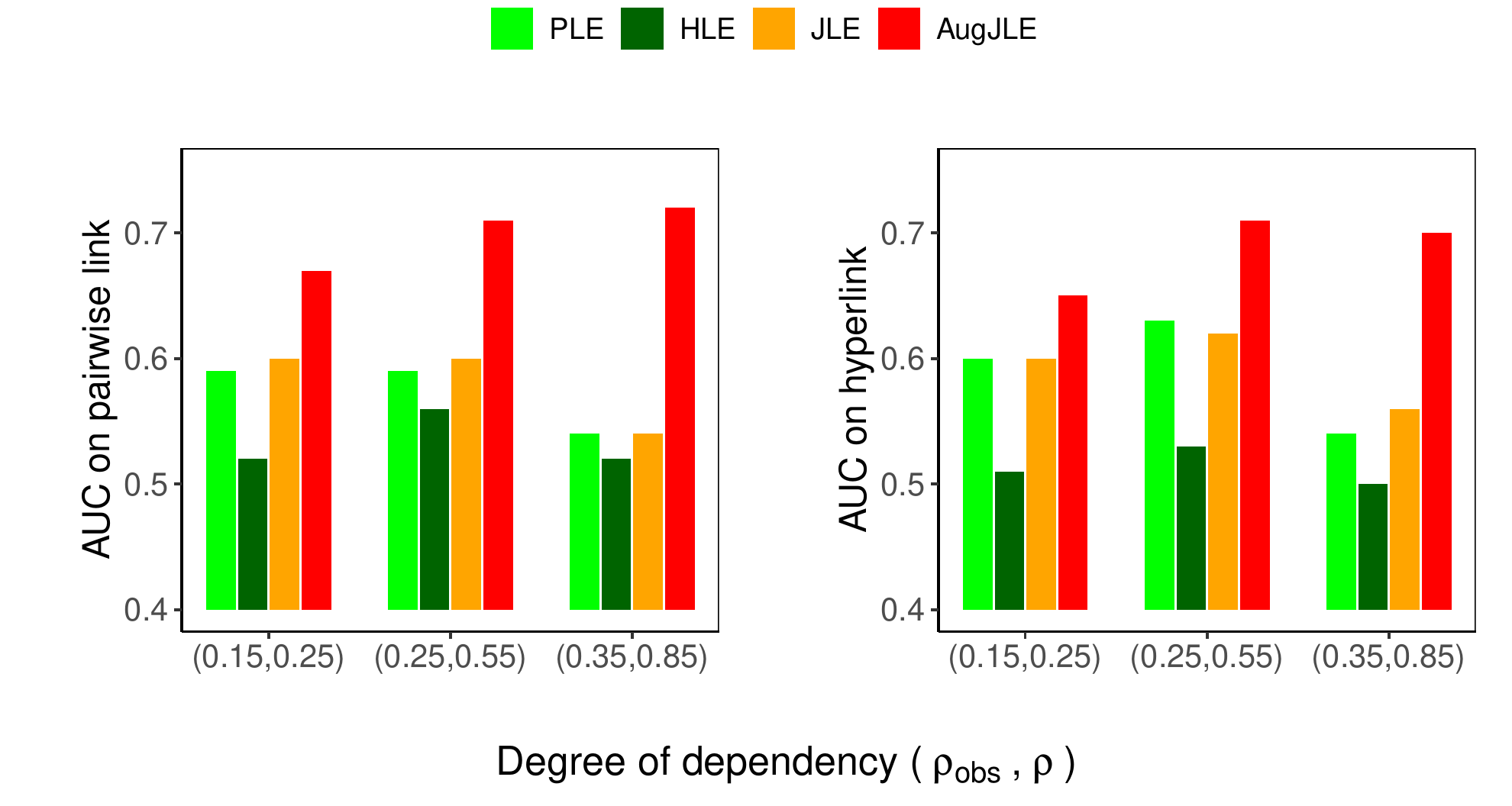}
 \label{fig_3}     				
      				\end{center}
      				
    \end{figure}

 
\begin{table}[h]
\centering
\caption{{\small The link predictions performance of different methods. {The observation dependency $\rho_{obs}$ is fixed}. The network size is $N = 120$.}}

\begin{adjustbox}{width=0.71\textwidth}
\begin{tabular}{| c | c | c | c | c |  } 
  \hline \hline

 $\rho_{obs}= 0.25,\; \rho = 0.25$ & PLE & HLE  & JLE &\textbf{Aug JLE}    \\ \hline

{AUC$_{pair}$} & 0.56(0.01) &  0.52(0.01) & 0.59(0.02)  &  \textcolor{red}{0.64(0.01)}  \\

{AUC$_{hyper}$} &  0.58(0.02) & 0.53(0.01) & 0.61(0.01)  & \textcolor{red}{0.66(0.02)}  \\
 \hline

  $\rho_{obs}= 0.25,\; \rho = 0.55$ & PLE & HLE & JLE &\textbf{Aug JLE}    \\ \hline

{AUC$_{pair}$} & 0.57(0.02) & 0.54 (0.01) & 0.60(0.01)  &  \textcolor{red}{0.67(0.01)}  \\

{AUC$_{hyper}$} &  0.59(0.02) & 0.55(0.01) & 0.60(0.01)  & \textcolor{red}{0.68(0.01)}  \\
\hline

$\rho_{obs}= 0.25,\; \rho = 0.85$ & PLE  & HLE & JLE &\textbf{Aug JLE}    \\ \hline

{AUC$_{pair}$} & 0.58(0.01) &  0.54(0.01) & 0.60(0.01)  &  \textcolor{red}{0.69(0.01)}  \\

{AUC$_{hyper}$} &  0.56(0.02) & 0.53(0.01)  & 0.59(0.02)  & \textcolor{red}{0.69(0.02)}  \\

\hline

  \hline
   \hline
\end{tabular}
\end{adjustbox}
\label{tab_3}
\end{table}

To better illustrate the strength of the hyperlink augmentation, we also investigate the performance of the proposed methods given the hyperlink generating model is misspecified. Similarly, {we consider a pairwise link network with $K=12$ clusters among $N$ nodes such that nodewise latent factors $\bm{Z}$ within the same cluster 
are closer. The pairwise links $\{Y_{ij}\}$ are independently generated based on the latent positions as in (\ref{link_form_1}).
Therefore, the probability of pairwise link within the same cluster is about 0.90 and is about 0.25 for between-cluster links. Based on $3$-order $\bigtriangleup_{i_1i_2i_3}$, we assume that a hyperlink status $Y_{i_1i_2i_3}$ is independently generated from the Bernoulli distribution with probability $P(Y_{i_1i_2i_3}=1)$ as: 
\vspace{-3mm}
\begin{align*}
{\small
P(Y_{i_1i_2i_3}=1)=
\begin{cases}
0.90,\; \text{if} \; i_1,i_2,i_3\; \text{are in same cluster and}\; \bigtriangleup_{i_1i_2i_3}=1  ;\\
0.10,\; \text{if} \; i_1,i_2,i_3\; \text{are not in same cluster and}\; \bigtriangleup_{i_1i_2i_3}=1;\\
0,\; \text{otherwise},
\end{cases} }
\end{align*}
and the link dependency can be quantified as $\rho: = corr\Big\{\mathbf{1}_{\{Y_{i_1i_2i_3}=1\}}(Y_{i_1i_2i_3}),\mathbf{1}_{\{\bigtriangleup_{i_1i_2i_3}=1\}}(\bigtriangleup_{i_1i_2i_3})\Big\}$. 
Intuitively, $\rho$ measures the marginal correlation between a hyperlink among $i_1,i_2,i_3$ and the corresponding clique among $i_1,i_2,i_3$. Under this generating model, the dependency between pairwise links and hyperlinks is $\rho = 0.8$. Notice that the generation of hyperlinks is not based on the latent factors which are used to formulate pairwise links. Instead, hyperlinks are generated upon clique $\bigtriangleup$ to preserve the dependency with pairwise links. 
Figure \ref{fig_1} shows that the augmented \text{JLE} has a significant improvement on both pairwise link prediction and hyperlink prediction compared with other methods given that hyperlink model is misspecified while the link dependency is strong.} Specifically, the improvement from the augmented \text{JLE} over \text{JLE} is $25\%$ on pairwise link prediction and $32\%$ in hyperlink prediction when the network size is $N=360$.

\vspace{-2mm}
\begin{figure}[h]
\caption{{\small The comparison of link prediction among augmented \text{JLE} (\text{Aug JLE}); JLE using observed network (\text{JLE}), link embedding using observed pairwise links (\text{PLE}), and using observed 
hyperlinks (\text{HLE}). The network size varies from $N = 120$ to $N=360$. The observation dependency $\rho_{obs}$ is 0.2. }}
\vspace{-8mm} 
       				\begin{center}
      					\includegraphics[width=5.4in,height=2.3in]{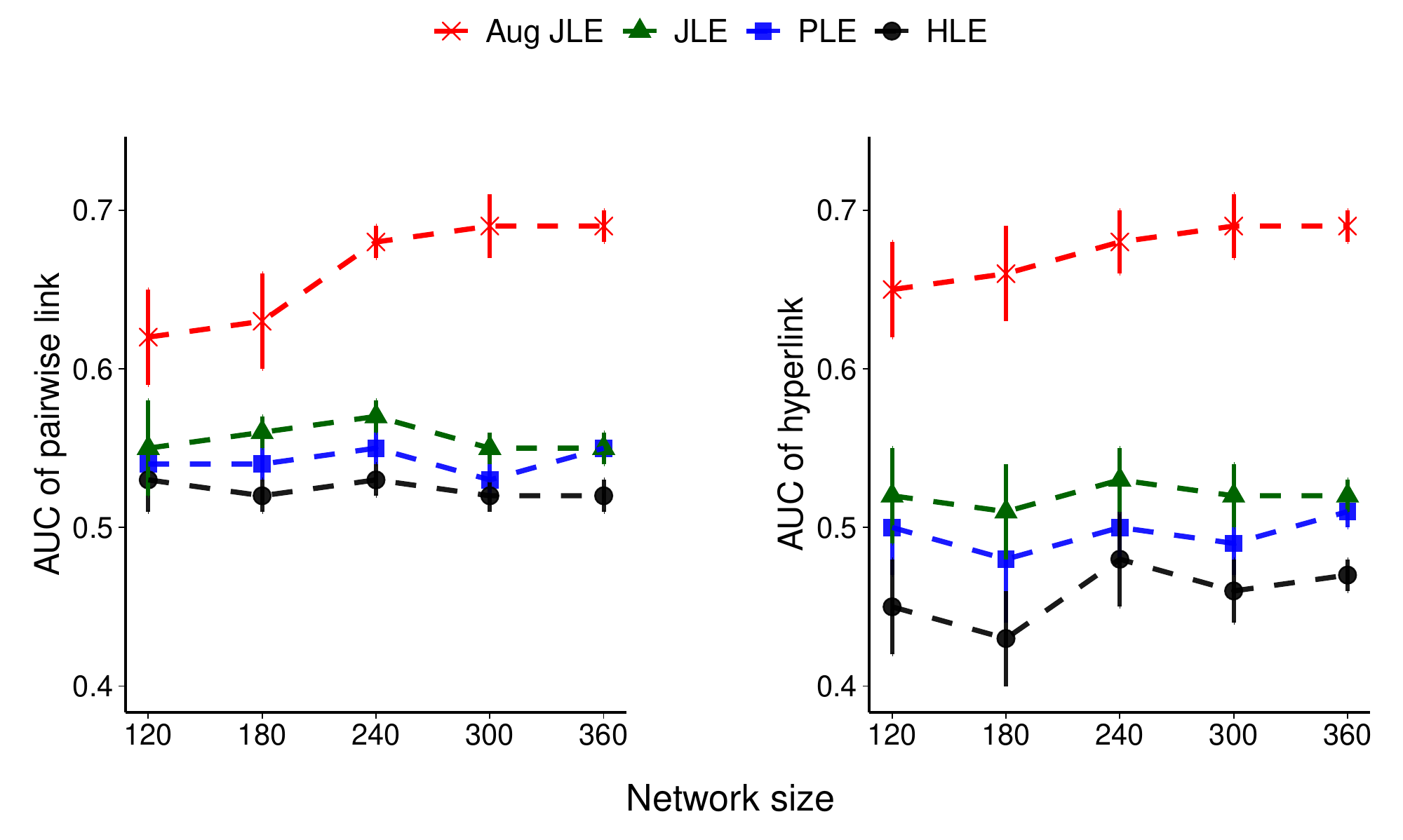}
 \label{fig_1}     				
      				\end{center}
      				
    \end{figure}

\vspace*{-7mm}
\section{Real Data Application}\label{2.5}

{In this section, we apply the proposed joint embedding method to the Facebook social circles network 
dataset \citep{leskovec2012learning}, 
which 
contains the 
ego-network \citep{kim2017community}. Each ego denotes a specific user in Facebook, and his or her associated ego-network is the social network corresponding to this user's friends in Facebook. One of the important attributes of ego-network is that it contains the social circles defined by the user, which leads to subgroups among people in the ego-network. Currently, the social circle is a common functionality for many popular social media such as Facebook, Google and Twitter. The purpose of introducing the social circle is to allow users to organize their own social network to mitigate the 'information overload' through filtering contents or status updates posted by friends in specific groups. In addition, it allows users to protect their privacy by restricting or sharing personal information for specific groups of friends.}  

{The major distinction between social circles and traditional social network communities is that social circles are in general highly overlapped and can be hierarchically nested, and therefore different people in an ego-network might belong to multiple circles. In addition, unlike the social network communities identified with dense internal connections, social circles are formulated through specific attributes of friends selected by the user. For example, a user might cluster his or her social networks according to categories such as college friends, high school friends, department friends or colleagues.} 

{Currently, social media adopts two methods to formulate social circles in an ego-network. The first one requires the user to manually group people, which is 
time-consuming and cannot be updated automatically when the user adds more friends. The second approach categorizes social circles through identifying people sharing common predefined attributes. However, it fails to incorporate the user's individual preferences and suffers 
from missing profile information. In this section, we apply the proposed approach and investigate the performance of learning the user-specified clustering through network embedding.} 

{The network data we consider here includes 224 nodes as users and 6384 undirected pairwise links as friend relationships. In addition, there are 14 overlapped circles within the network, which are formulated according to the similarity of social relation features among people defined by the users, such as  common university affiliations, sports teams and relatives. These circles usually contain a large number of nodes so that the average size of each circle is 40 nodes and the largest circle contains 201 nodes. The ego-network is illustrated in Figure \ref{fig:Fig4}.} 

We first randomly split the pairwise links from the ego-network into training, validation and test sets with the proportion of $40\%, 20\%$ and $40\%$.  
Incorporating multi-way information requires more elaborate preprocessing. We discard the most non-informative social circles, i.e., the largest social circle with a size of 201, since the large social circle is non-informative in differentiating subgroup-specific attributes, and denote the 13 remaining social circles as $\{C_l\}, l = 1,\cdots,13$.
We then decompose each social circle into a set of three-way hyperlinks $\Omega_{\mathcal{Y}}^{pool} =  \{Y_{ijk}\}$ according to the following rule:
\begin{align}\label{social_circle}
Y_{ijk} =  \begin{cases} 1, \;|\{l|i\in C_l, j\in C_l, k\in C_l, l = 1,\cdots, 13 \} | \geq 1 \\
0, \; |\{l|i\in C_l, j\in C_l, k\in C_l, l = 1,\cdots, 13 \} | = 0
\end{cases},
\end{align}
where $|\cdot|$ denotes cardinality of a set. Through this preprocessing, the local multi-way relations and the global subgroups in the original social circles can be represented by the three-way relations constructed in (\ref{social_circle}). {Finally, we generate training hyperlink statuses $\Omega_{\mathcal{Y}}$ through randomly sampling $1\%$ observed hyperlink statuses from $\Omega_{\mathcal{Y}}^{pool}$. The size of $\Omega_{\mathcal{Y}}$ is about 600 after the preprocessing, and the proportions for present links $\{Y_{ijk}\in \Omega_{\mathcal{Y}}|Y_{ijk}=1\}$ and absent links $\{Y_{ijk}\in \Omega_{\mathcal{Y}}|Y_{ijk}=0\}$ are relatively balanced. For applying the augmented JLE, we also generate augmented hyperlink statuses via utilizing the pairwise friendships $\mathbf{Y} =  \{Y_{ij}\in \{0,1\}\}_{1\leq i \neq j \leq 224}$. Specifically, we collect the candidate pool for augmented hyperlink statuses $\Omega_{pool} = \{\hat{Y}_{ijk}\}$ as follows:
\begin{align*}
\hat{Y}_{ijk} =  \begin{cases} 1, \; Y_{ij}=Y_{ik}=Y_{jk}=1,\;\text{and $(i,j,k)$ in the same circle},\\ 
0, \; Y_{ij}=Y_{ik}=Y_{jk}=0,\;\text{and $(i,j,k)$ not in the same circle}.
\end{cases}
\end{align*} 
Then we construct the augmented hyperlink statuses set $\hat{\Omega}_{\mathcal{Y}}$ via randomly sampling 2000 augmented hyperlink statuses from $\Omega_{pool}$ with a balanced proportion between presence and absence of links.}

\begin{figure}[h]
\centering
   \includegraphics[width=4in,height=2.2in]{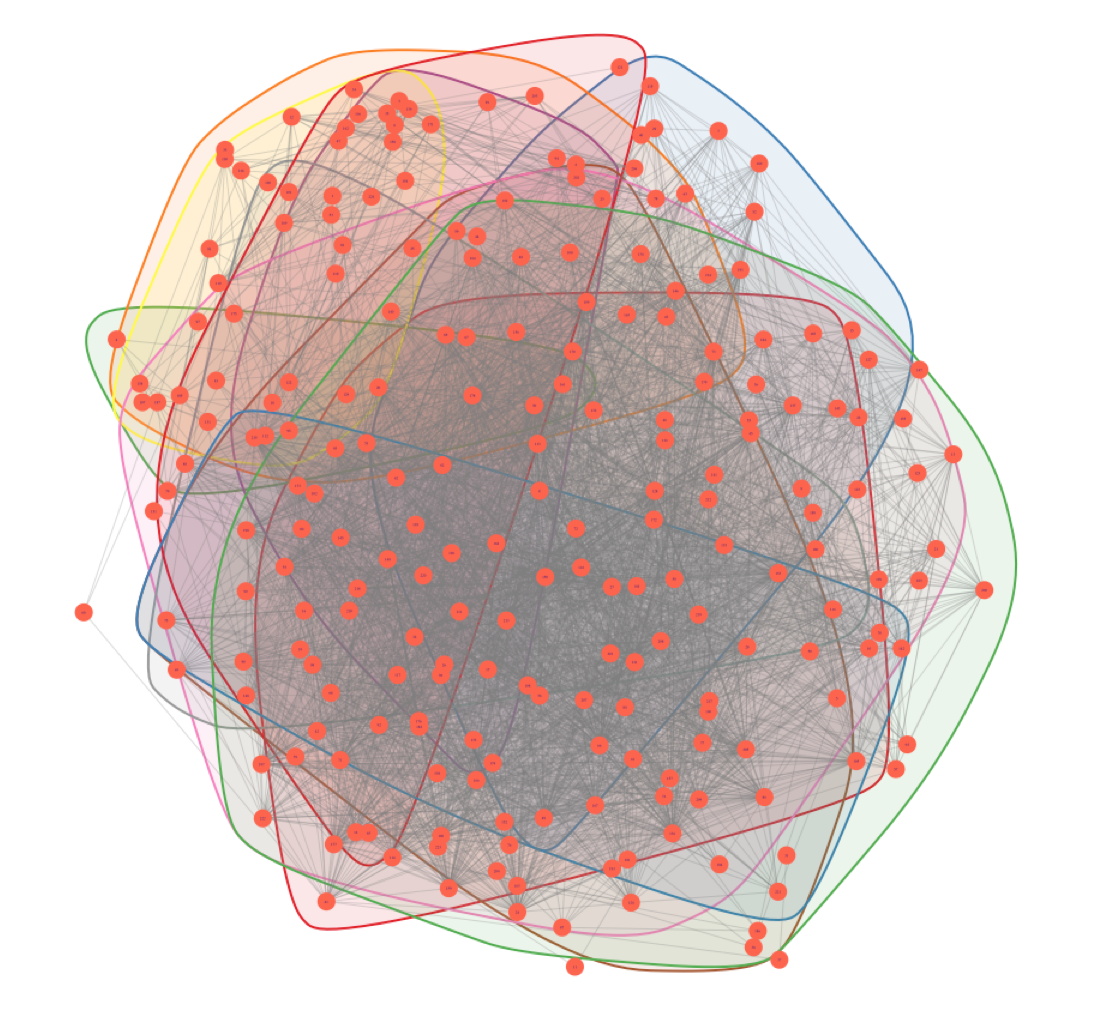}
\caption{The ego network in the Facebook dataset,  where the social circles are marked as polygons with different colors}
\label{fig:Fig4}
\end{figure}    


We investigate the performance of the proposed method and other competing network embedding methods on predicting both two-way relations on the ego-network.
In addition, we also investigate the performance of hyperlink prediction to evaluate whether the social circle information has been encoded into nodes' latent factors. Instead of directly predicting the original social circles, we predict the joint memberships of specific $m$ nodes, i.e., whether they belong to the same social circle or not, and compare the result with the original circles. 
We choose the order of testing hyperlinks as $m=6, 10$, and the prediction is based on estimated $\bm{Z}$ through
$P(Y_{i_1i_2\cdots i_m}=1) = \frac{\exp(\sum_{(i,j)\in \{1,\cdots, m\}}Z_iZ_j^T)}{1+\exp(\sum_{(i,j)\in \{1,\cdots, m\}}Z_iZ_j^T)}$. 
For the proposed joint embedding method JLE, and the counter-method of only using PLE, the rank of $\bm{Z}$ is chosen at $r=5$. For the tuning parameters of competing methods, we adopt the strategies recommended by the original papers or the packages. 

\begin{table}[h]
\centering
\caption{AUC of link prediction for ego-network}
\vspace{-3mm}
\resizebox{12cm}{!}{
\begin{tabular}{| c | c c | c | c | } 
   \hline
   \hline
   
   & \multicolumn{4}{c|}{\textbf{Link Prediction}} \\
   \hline
  &  \multicolumn{2}{c | }{\textbf{Pairwise Link}}    &  \multicolumn{1}{c | }{\textbf{6-order Hyperlink} } &  \multicolumn{1}{c | }{\textbf{10-order Hyperlink} }   \\ 
\cline{2-5}  
   
  & test & global & test &test  \\   \hline 

 \textbf{Aug JLE}  & \textcolor{red}{0.80} & \textcolor{black}{0.82} & \textcolor{red}{0.95} &\textcolor{red}{0.97}  \\  
  
   \textbf{JLE}  & {0.79} & \textcolor{black}{0.82} & {0.91} &{0.89}  \\
 HLE & 0.57 &  0.56 & 0.89& 0.82   \\
   PLE & 0.80 &  \textcolor{red}{0.83} & 0.62 & 0.63  \\
   GraRep & 0.79 &  0.81 & 0.77 & 0.51   \\
   LINE & 0.62 &  0.59 & 0.51 & 0.75   \\
   Node2Vec & 0.49 &  0.49 & 0.49 & 0.48  \\
   \hline

  \hline
   \hline
\end{tabular}}
\label{tab6}
\end{table} 

%
%
%
%
%
{The comparison of performance is illustrated in Table \ref{tab6}. For the two-way relation prediction, the proposed JLE and augmented JLE method achieve top performances compared with competing methods. It achieves about $30\%$ improvement over the \text{LINE} and $60\%$ improvement over the \text{Node2Vec}. The real data analyses show the importance of borrowing  high-order relation information for the two-way relation predictions. The incorporated third-order hyperlinks recover the underlying subgroup structure, and the estimated latent features characterize the subgroup attributes. If partial subgroup information is recovered, the two-way relations from the subgroup level provide additional information regarding potential friendships. The inferior performance of LINE and Node2Vec might be due to 
the relatively sparse ego-network as the pairwise links only account for $13\%$ of the total possible friendships.}
 
{In terms of the multi-way relation prediction, the proposed augmented JLE method achieves the best performance for both 
6-order and 10-order hyperlink prediction. Specifically, the Aug JLE achieves $23\%$ improvement over the best existing method Grarep on the 6-order hyperlink prediction and $29\%$ improvement over the best existing method LINE on the 10-order hyperlink prediction. Although the partial embedding method PLE and the GraRep perform well for the two-way relation prediction, they do not have consistent performance for hyperlink predictions, indicating that the social circles indeed encode significant high-order relations information which cannot be represented through two-way relations. This demonstrates the importance of incorporating high-order information for predicting multi-way relations. 

\color{black}

\section{Discussion} 
In this paper, we propose a new network link prediction method. The major innovation of the proposed method is to incorporate the multi-way relation into the network embedding process and therefore jointly embed both the hyperlinks and pairwise links. It allows the node-wise latent factors to encode the pairwise similarity to their neighbourhood, and induce cohesive high-order subgroups. In addition, the proposed method 
formulates hierarchical modeling for the 
link-generating process to introduce the dependency between pairwise links and hyperlinks. In terms of estimating latent factors, the link dependency allows borrowing the mutual information between pairwise links and hyperlinks such that prediction for both pairwise links  and hyperlinks  can be improved. In term of model interpretability, the link dependency reflects the principle that high-order interaction among nodes in a network can in general be built on the low-order interactions. {In theory, we establish the consistency of the link probability estimator based on the proposed joint embedding loss function. In addition, we show that the convergence rate can be improved through incorporating the observed hyperlinks and hyperlink augmentation. 



In this paper, we only consider a latent-factor based hyperlink augmentation approach. {However, it is worth further exploration to develop an inference procedure to learn the hierarchical dependency of hyperlinks on pairwise links.} {In addition, generalization of the proposed method to varying hyperlink orders is also one of the important directions for our future study.}

\section*{Acknowledgements} The authors thank the Associate Editor and two anonymous reviewers for their suggestions and helpful feedback which improved the paper significantly. 
\vspace*{-3mm}
\section*{Supplementary Materials}
The supplementary materials provide proofs of the Theorem \ref{theorem1}, Theorem \ref{theorem2}, optimization algorithm in Section 3.5, explicit formulation for hyperlink modeling, detailed discussion on the theoretical results, detailed settings and results of simulations, and ROC graphs. 
\vspace*{-3mm}
{\footnotesize\bibliography{unblinded_minor_revision_main_text}}

\end{document}